\documentclass[preprint,12pt]{elsarticle}

\usepackage{natbib}
\usepackage{amssymb,amsfonts,dsfont,float,graphics,epsfig,epstopdf,color,verbatim,tabularx,bm,multirow}
\usepackage{ulem}

\usepackage{subfigure}
\usepackage{amsmath}
\usepackage{caption,graphicx}
\usepackage{soul}
\usepackage{color,xcolor}

\usepackage[colorlinks,
            linkcolor=blue,
            urlcolor=blue,
            citecolor=blue]{hyperref}
\journal{}

\begin{document}

\begin{frontmatter}


\title{Probability distributions for kinetic roughening in the Kardar-Parisi-Zhang growth with long-range spatiotemporal correlations}


\author{Zhichao Chang}
\author{Hui Xia\corref{cor1}}
\cortext[cor1]{corresponding author}
\address {School of Materials Science and Physics, China University of Mining and Technology, Xuzhou 221116, China}
\ead{hxia@cumt.edu.cn}
\begin{abstract}
We investigate numerically the effects of long-range temporal and spatial correlations based on {the rescaled distributions of  the squared interface width $W^2(L,t)$ and the interface height $h(x,t)$ in the (1+1)-dimensional Kardar-Parisi-Zhang (KPZ) growth system within the early growth regimes. Through extensive numerical simulations, we find that long-range temporally correlated noise could not significantly impact the distribution form of the interface width. Generally, $W^2(L,t)$ obeys approximately lognormal distribution  when the temporal correlation exponent $\theta \ge 0$.
 On the other hand, the effects of long-range spatially correlated noise are evidently different from the temporally correlated case. Our results show that, when the spatial correlation exponent $\rho \le 0.20$, the distribution forms of  $W^2(L,t)$ approach the lognormal distribution, and when $\rho > 0.20$, the distribution becomes more asymmetric, steep, and fat-tailed, and tends to an unknown distribution form. As a comparison, probability distributions of the interface height are also provided in the temporally and spatially correlated KPZ system, exhibiting quite different characteristics from each other within the whole correlated regimes. For the temporal correlation, the height distributions satisfy Tracy-Widom Gaussian orthogonal ensemble (TW-GOE) when  $\theta \to 0$, and with increasing $\theta$, the height distributions crossover continously to an unknown distribution. However, for the spatial correlation, the  height distributions gradually transition from the TW-GOE  distribution to the standard Gaussian form.}
\end{abstract}




\end{frontmatter}


\section{Introduction}
\label{}
Nonequilibrium processes are common in nature, which play indispensable roles in many actual physical phenomena, such as kinetic roughening, turbulence, combustion and active matter, among others
\cite{Barabasi1995,Chen2016,Veynante2002}. As a typical nonequilibrium system, kinetic roughening has received much attention for several decades\cite{Barabasi1995,Kardar1986,Family1985,Chu2016,Halpin1995}. The Kardar-Parisi-Zhang (KPZ) equation, which was originally used to describe kinetic roughening process\cite{Kardar1986}, is one of the most important nonlinear Langevin-type growth equations in this field, and it can also be observed in many other fields\cite{Barabasi1995,Squizzato2018,Kulkarni2013,Gladilin2014,Altman2015,Mathey2015,Chen2016}. The KPZ equation in the (1+1)-dimensions reads\cite{Kardar1986},
\begin{equation}
    \frac{{\partial {{h}}\left( {x,t} \right)}}{{\partial t}} = \nu {\nabla ^2}h + \frac{\lambda }{2}{\left( {\nabla h} \right)^2} + \eta \left( {x,t} \right),
\end{equation}
where $h(x,t)$ is the interface height in position $x$ and time $t$, $\nu$ is the surface tension describing relaxation of the interface, $\lambda$ is the nonlinear coefficient, representing lateral growth, and $\eta(x,t)$ is stochastic force that usually is uncorrelated
\begin{equation}
    \left\langle {\eta \left( {x,t} \right)\eta \left( {x',t'} \right)} \right\rangle  \sim \delta \left( {x - x'} \right)\delta \left( {t - t'} \right).
    \label{eq2}
\end{equation}

Previous research has indicated that surface roughening processes exhibit scaling behaviors\cite{Family1985,Zhang1990,Hentschel1991,Hanfei1993}, which are often characterized by the fluctuation of surface and interface height. To describe the characteristic of kinetic roughening, {one often uses the squared interface width that is defined by\cite{Barabasi1995}
\begin{equation}
   {W^2}(L,t) = \left\langle {\sqrt {\frac{1}{L}\sum\limits_x {{{\left[ {h(x,t) - \bar h(t)} \right]}^2}} }}\right\rangle,
\end{equation}
where ${\bar h\left( t \right)}$ is the mean height of rough interface with substrate size $L$, $\left\langle  \cdots  \right\rangle $ stands for the average over different independent noise. And $W^2(L,t)$ follows a dynamic scaling form\cite{Family1985}:
\begin{equation}
    W^2\left( {L,t} \right) \sim \left\{ {\begin{array}{*{20}{c}}
{{t^{2\beta} }, \ {{ for \ t}} \ll {{{t}}_\times},}\\
{{L^{2\alpha} }, \ {{ for \ t}} \gg {{{t}}_\times},}
\end{array}} \right.
\end{equation}
where $\beta$ is the growth exponent, $\alpha$ is the roughness exponent, and the characteristic time ${t_\times} \sim {L^z}$ with the dynamic exponent $z = {\alpha / \beta }$.}  These scaling exponents could reveal plenty of dynamic scaling properties of surface and interface growth\cite{Barabasi1995}.

The (1+1)-dimensional KPZ equation with Gaussian noise has well studied in recent decades, the scaling exponents and universal distributions are also obtained exactly\cite{Spohn2000,Sasamoto2010,Takeuchi2013,Takeuchi2018}. However, the KPZ equation driven by long-range temporally or spatially correlated noise is more complicated and the exact solutions are still lacking. Fortunately, based on numerical simulations and analytic approximate techniques, for example, dynamic renormalization group (DRG)\cite{Medina1989} and self-consistent expansion (SCE)\cite{Katzav1999,Katzav2004}, a lot of rich analytical predictions and numerical results have been achieved. But, generally speaking, the discrepancy among these theoretical predictions and numerical solutions is still very obvious\cite{Medina1989,Katzav1999,Katzav2004,Song2021}.
Previous research mainly focused on the stochastic growth systems with long-range spatial correlations rather than those with long-range temporal correlations \cite{Katzav1999,Meakin1989,Meakin1990,Amar1991,Peng1991,Pang1995,Mukherji1997,Li1997,Chattopadhyay1998,Frey1999,Verma2000,Katzav2002a,Katzav2003,Katzav2003b,Katzav2006,Katzav2013,Kloss2014,Chu2016,Morais2011,Xia2016,Niggemann2018,Fedorenko2008}. Although recent studies about the temporally correlated KPZ equation filled partial research gaps\cite{Lam1992,Katzav2004,Strack2015,Song2016,Squizzato2019,Lopez2019,Xia2020, Song2021,Song2021b,Hu2023}, some open issues need further clarification.
It is generally agreed that long-rang correlations could affect scaling properties of KPZ growth\cite{Barabasi1995,Halpin1995}, while some disagreements still exist from the perspective of the estimated values of critical exponents based on analytical predictions and numerical simulations\cite{Medina1989,Hanfei1993,Katzav1999,Katzav2004,Fedorenko2008,Lopez2019,Song2021b,Hu2023}.

{The probability distributions of both the interface height $h(x,t)$ and the squared interface width  $W^2(L,t)$ could display plenty of universal features and, accordingly, various continuum growth equation and discrete growth models have been analyzed and distinguished in terms of universality classes.
It is now well-known that, mainly due to excellent theoretical calculations\cite{Spohn2000,Sasamoto2010,Takeuchi2013}, the distributions of interface height in the (1+1)-dimensional KPZ are characterized by Tracy-Widom (TW) distributions\cite{Tracy1994} during the early growth regimes, whereas the stationary height distributions are depicted by Baik-Rains (BR) form\cite{Baik2000} at the saturated regimes.} 


Previous investigations demonstrated that { the probability distributions of the squared interface width,  $P(W^2)$ at early growth and steady state regimes exhibits various statistical characteristics \cite{Racz1994, Foltin1994, Antal1996, Marinari2002, Aarao2005, Paiva2007, Oliveira2007, Miranda2008, Kelling2011, Almeida2014, Halpin-Healy2014, Halpin-Healy2015, Carrasco2016, Almeida2017}. Usually, there exist two relevant types of $P(W^2)$ in order to uncover universal behavior of the underlying growth dynamics: one type is the interface width distributions measured for stochastic growth system in the steady state regime with periodic boundary conditions \cite{Racz1994, Foltin1994, Antal1996, Marinari2002, Aarao2005, Oliveira2007, Miranda2008, Kelling2011, Halpin-Healy2015}, and another type is the the interface width distributions measured at the early growth regime with window boundary conditions \cite{Paiva2007, Almeida2014, Halpin-Healy2014, Carrasco2016, Almeida2017}.} As a special case, the early-time limits of linear and nonlinear stochastic growth are also interesting since the squared interface width distributions are found to closely approximate the lognormal distribution\cite{Antal1996, Carrasco2016}.
These related numerical researches mentioned above focus mainly on the local linear and nonlinear growth system, yet when long-range temporal and spatial correlations are introduced, the evolution of interface height and width displays nontrivial nonlocal characteristics. {To the best of our knowledge, the theoretical investigations on interface width and height distributions remain inadequate  in the literature  for this type of interface growth in the presence of long-range correlations.
Therefore, these motivate us to investigate the effects of long-range temporal and spatial correlations on the continuous KPZ system on the basis of the interface width and height  distributions, that is,  $P(W^2)$  and $P(h)$}, which allow one to further obtain some important characteristics for kinetic roughening.

Considering that introducing long-range correlated noise into these KPZ growth system needs enormous and unbearable computational time which limits our current investigations on the steady state width distribution, thus, 
in this work we focus mainly on { the rescaled distributions of both the interface width and height in the temporally and spatially correlated KPZ growth systems at the early growth times.
 Firstly, we will re-examine scaling properties of the KPZ system driven by long-range correlated noises, and then investigate if and how long-range temporal and spatial correlations impact  $P(W^2)$ and $P(h)$} within early growth regimes, and what are characteristics and differences of probability distributions in the KPZ growth driven by these two long-range correlated noises. As two special cases, when long-range temporal and spatial correlations are ignored, our results will be reduced to that of the normal KPZ with Gaussian white noise.

The paper is organized as follows: Firstly, we introduce the method to generate long-range correlated noises. And then, we describe one of the improved versions of the finite-difference (FD) method for direct simulating the KPZ equation driven by long-range temporally and spatially correlated noises. Next, we exhibit our numerical simulations and compare them with previous research. Finally, the corresponding discussions and conclusions are given.

\section{Basic methods and concepts}

\subsection{Generating long-range correlated noises}

When the noises have long-range temporal and spatial correlations, these two $\delta$ functions of Eq.\ref{eq2} are replaced by the term that decays as a power of time and distance. Thus, the second moment of the correlated noise is given by
\begin{equation}
    \left\langle {\eta \left( {x,t} \right)\eta \left( {x',t'} \right)} \right\rangle  \sim {\left| {x - x'} \right|^{2\rho  - 1}}{\left| {t - t'} \right|^{2\theta  - 1}},
\end{equation}
where $\rho$ and $\theta$ are the spatial and temporal correlation exponents, respectively. If $\rho = 0$ and $\theta \ne 0$, the noise is long-range temporally correlated case, on the contrary, the noise has long-range correlations in space.

To generate long series of correlated noise, we adopt the fast fractional Gaussian noise (FFGN) method, which was first proposed by Mandelbrot\cite{Mandelbrot1971}. One starts by generating a series of uncorrelated uniformly distribution numbers $\xi \left( u \right)$ in the range [0, 1]. The weight function is given by
\begin{equation}
    \omega_n^2 = \frac{{12\left( {1 - r_n^2} \right)\left( {{B^{\frac{1}{2} - \varphi }} - {B^{\varphi  - \frac{1}{2}}}} \right){{\left( {a{B^{ - n}}} \right)}^{1 - 2\varphi }}}}{{\Gamma \left( {2 - 2\varphi } \right)}},
\end{equation}
where ${r_n} = {e^{ - {u_n}}}$ with ${u_n} = a{B^{ - n}}$, $B = 2$, and $a = 6$. $\Gamma \left(  \cdots  \right)$ is Gamma function, and $\varphi $ represents temporal or spatial correlation exponent. And then, two autocorrelation decaying functions are defined by
\begin{equation}
    \begin{array}{l}
{X_1}\left( u \right) = \left[ {{\xi _1}\left( u \right) - 0.5} \right]/\sqrt {1 - {r^2}} , \ \ \ \ \ for \ t = 1,\\
{X_n}\left( u \right) = r{X_{t - 1}}\left( u \right) + \left[ {{\xi _t}\left( u \right) - 0.5} \right], \ for \ t > 1.
\end{array}
\end{equation}
Finally, one can obtain the long-range correlated noise
\begin{equation}
    {\eta} = \sum\limits_{n = 1}^N {{\omega_n}{X_t}\left( {{u_n}} \right)} ,
\end{equation}
where $N$ is the number of components needed, which should be increased to obtain the desired power-law exponent $\varphi$ with higher precision at low frequencies.

\subsection{The discretized schemes of (1+1)-dimensional KPZ equation}

FD method is one of the most direct and common numerical tools. Theoretically, the differential interval is enough small, and the numerical results one obtained will be more accurate\cite{Moser1991}. Unfortunately, numerical divergence in simulating the nonlinear KPZ system could not be avoided based on the standard FD method. In order to suppress the annoying growth instability, an exponentially decaying function was suggested to replace the nonlinear term, which could be partially effective at suppressing numerical instability\cite{Dasgupta1997,Miranda2008}. However, this exponentially decaying technique including infinitely many higher-order nonlinearities may cause nontrivial scaling behavior\cite{Gallego2016,Li2021}. Interestingly, an improved FD method proposed by Lam and Shin (LS)\cite{Lam1998} could suppress effectively numerical divergence in comparison with the normal FD scheme. Thus, the discretized KPZ equation with LS scheme in the (1+1)-dimensions has the following form,
\begin{equation}
{h}\left( {x,t + 1} \right) =h\left( {x,t} \right) + \Delta t\left[ {\nu \Phi \left( {x,t} \right) + \frac{\lambda }{2}\Psi \left( {x,t} \right) + \eta \left( {x,t} \right)} \right].
\end{equation}
Here, the discretized diffusive term ${\Phi \left( {x,t} \right)}$ reads
\begin{equation}
\Phi \left( {x,t} \right) = \left[ h\left( {x + 1,t} \right) - 2h\left( {x,t} \right) + h\left( {x - 1,t} \right)\right]/\Delta x^2,
\end{equation}
and the nonlinear term ${\Psi \left( {x,t} \right)}$ is discretized as
\begin{align}
\Psi \left( {x,t} \right) = & \frac{1}{3}\left\{ {{{\left[ {h\left( {x + 1,t} \right) - h\left( {x,t} \right)} \right]}^2} + {{\left[ {h\left( {x,t} \right) - h\left( {x - 1,t} \right)} \right]}^2}} \right. \nonumber \\
&\left. {+ \left[ {h\left( {x + 1,t} \right) - h\left( {x,t} \right)} \right]\left[ {h\left( {x,t} \right) - h\left( {x - 1,t} \right)} \right]} \right\}/\Delta {x^2}.
\end{align}

In our following simulations, the time evolution of interface starts from an initially flat $h(x,0)=0$ with periodic boundary conditions. We set $\nu = 1$, $\Delta t = 0.05$, $\Delta x = 1$, and $\eta \left( {x,t} \right)$ is long-range temporally or spatially correlated noise generated by FFGN. And then, we need to adjust $\lambda$ for a given temporal or spatial correlation exponent in order to ensure into the true KPZ scaling regime before numerical divergence appearing. 

\subsection{Skewness and kurtosis of statistical distribution}

 For a series of random variables $X_i$ that obey a certain distribution, skewness ($S$) could describe asymmetry of the distribution and is defined by
\begin{equation}
    S=\frac{1}{n}\sum_{i=1}^n{\left[ \left( \frac{X_i-\mu}{\sigma} \right) ^3 \right]},
\end{equation}
where $\mu$ is the mean value and $\sigma$ is the standard deviation. $S = 0$ means the distribution is symmetric, $S > 0$ is called as the positive skewness representing the distribution incline to left, on the contrary, $S < 0$ is negative skewness, and the distribution inclines to right.

Kurtosis ($K$) is another important statistic, which could describe the steepness of distribution, 
\begin{equation}
    K=\frac{1}{n}\sum_{i=1}^n{\left[ \left( \frac{X_i-\mu}{\sigma} \right) ^4 \right]}.
\end{equation}
$K = 3$ represents the normal distribution, $K < 3$ means the distribution is lower and fatter than the normal distribution, and $K > 3$ indicates the distribution exhibits taller and thinner than the normal one.
According to measuring $S$ and $K$, one could compare a given distribution from a normal case and then estimate its specific form.

\section{Numerical results and discussions} 

\subsection{The KPZ equation with long-range temporal correlation}

\begin{figure}[htbp]
\centering
\subfigure{
\begin{minipage}[t]{0.40\linewidth}
\centering
\includegraphics[width=2.3in]{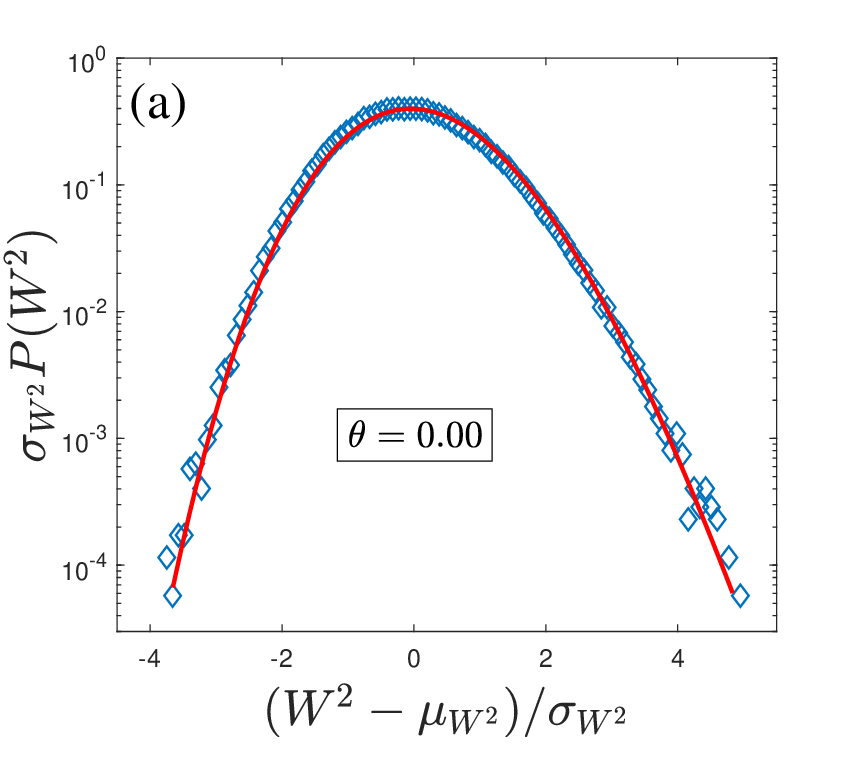}
\end{minipage}
}
\subfigure{
\begin{minipage}[t]{0.40\linewidth}
\centering
\includegraphics[width=2.3in]{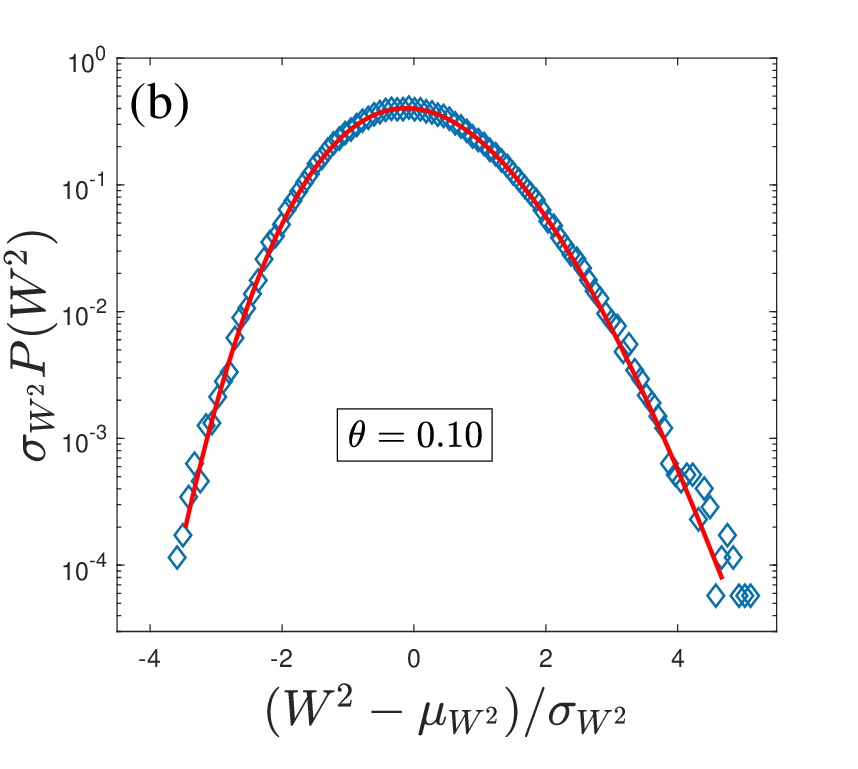}
\end{minipage}
}

\subfigure{
\begin{minipage}[t]{0.40\linewidth}
\centering
\includegraphics[width=2.3in]{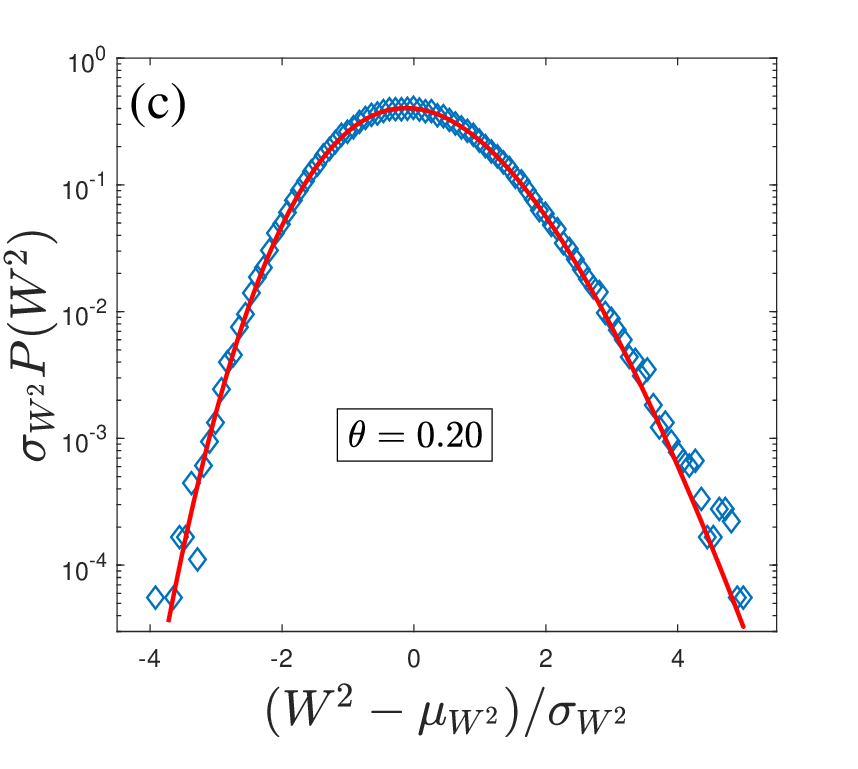}
\end{minipage}
}
\subfigure{
\begin{minipage}[t]{0.40\linewidth}
\centering
\includegraphics[width=2.3in]{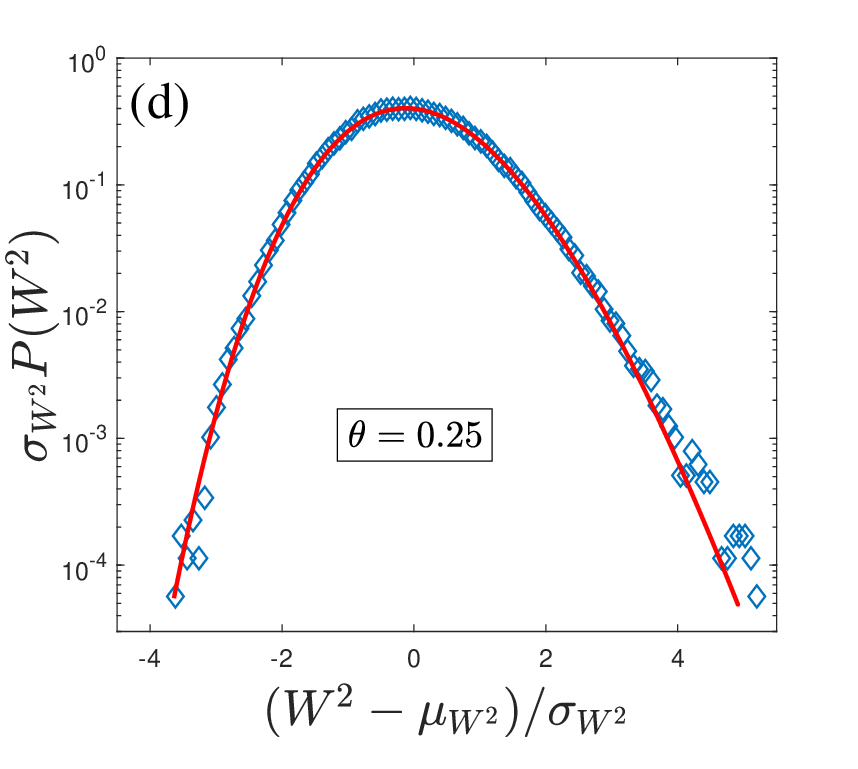}
\end{minipage}
}

\subfigure{
\begin{minipage}[t]{0.40\linewidth}
\centering
\includegraphics[width=2.3in]{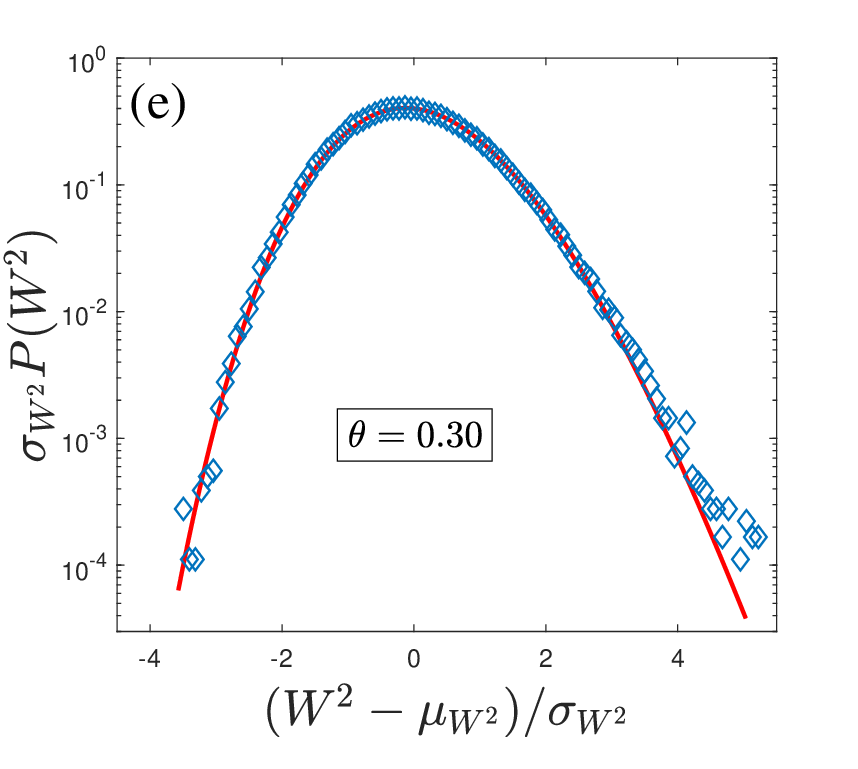}
\end{minipage}
}
\subfigure{
\begin{minipage}[t]{0.40\linewidth}
\centering
\includegraphics[width=2.3in]{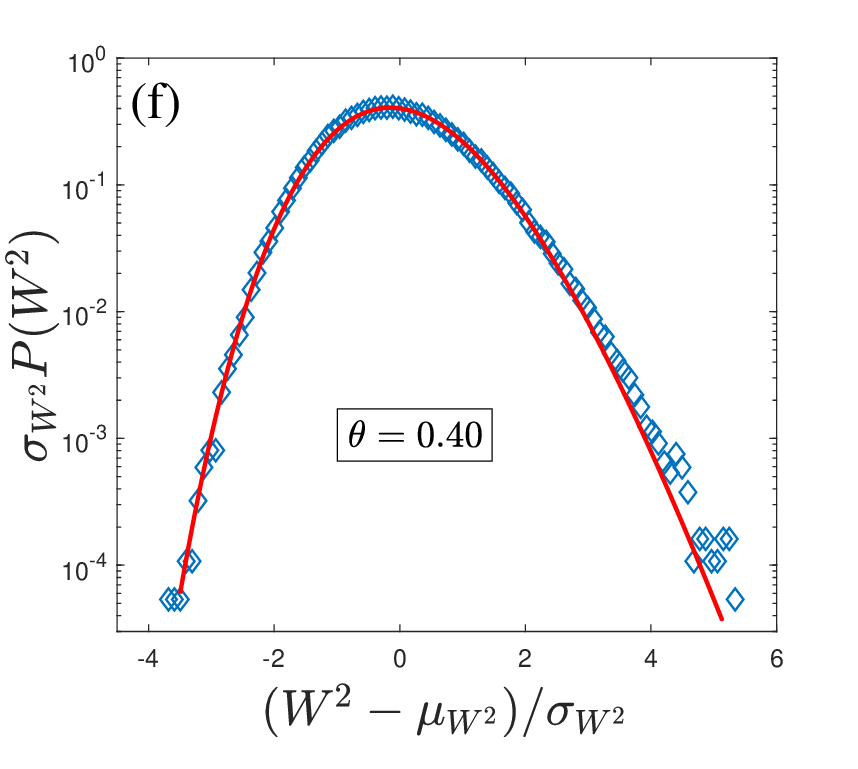}
\end{minipage}
}
\caption{{The rescaled distributions for $W^2(L,t)$ in the temporally correlated KPZ equation with different $\theta$: (a) $\theta = 0.00$, (b) $\theta = 0.10$, (c) $\theta = 0.20$, (d) $\theta = 0.25$, (e) $\theta = 0.30$, (f) $\theta = 0.40$. Here, $\mu_{W^2}$ and $\sigma_{W^2}$ are the mean value and standard deviation of $W^2(L,t)$, respectively. All data are averaged over $10^5$ independent realizations.  Comparison with lognormal distribution (red solid line) is provided correspondingly.}}
\label{Fig1}
\end{figure}

\begin{figure}[htbp]
\centering
\subfigure{
\begin{minipage}[t]{0.40\linewidth}
\centering
\includegraphics[width=2.3in]{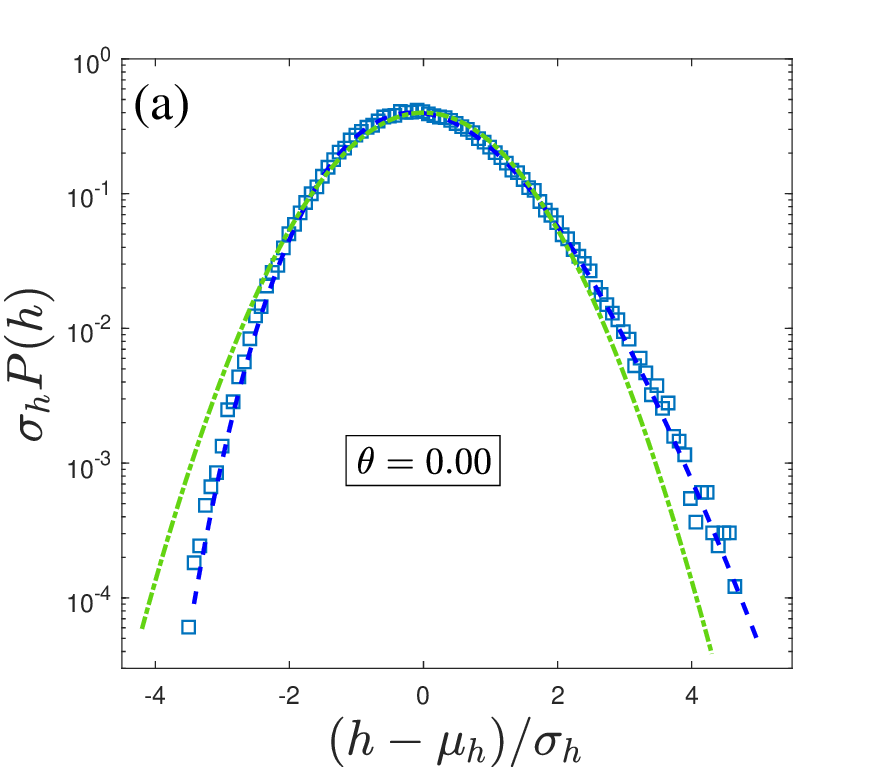}
\end{minipage}
}
\subfigure{
\begin{minipage}[t]{0.40\linewidth}
\centering
\includegraphics[width=2.3in]{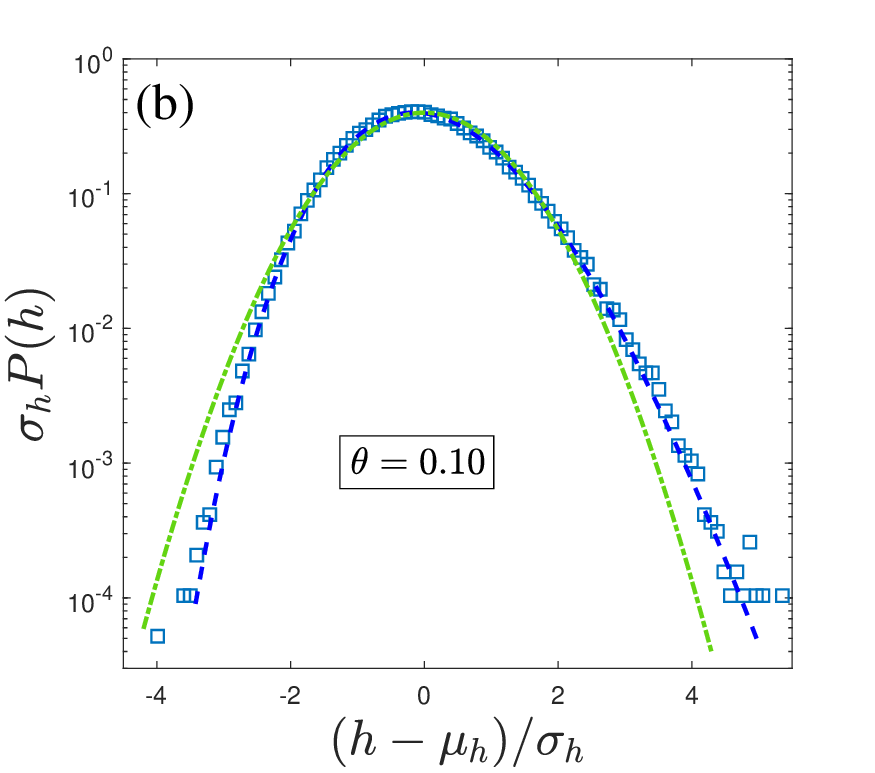}
\end{minipage}
}

\subfigure{
\begin{minipage}[t]{0.40\linewidth}
\centering
\includegraphics[width=2.3in]{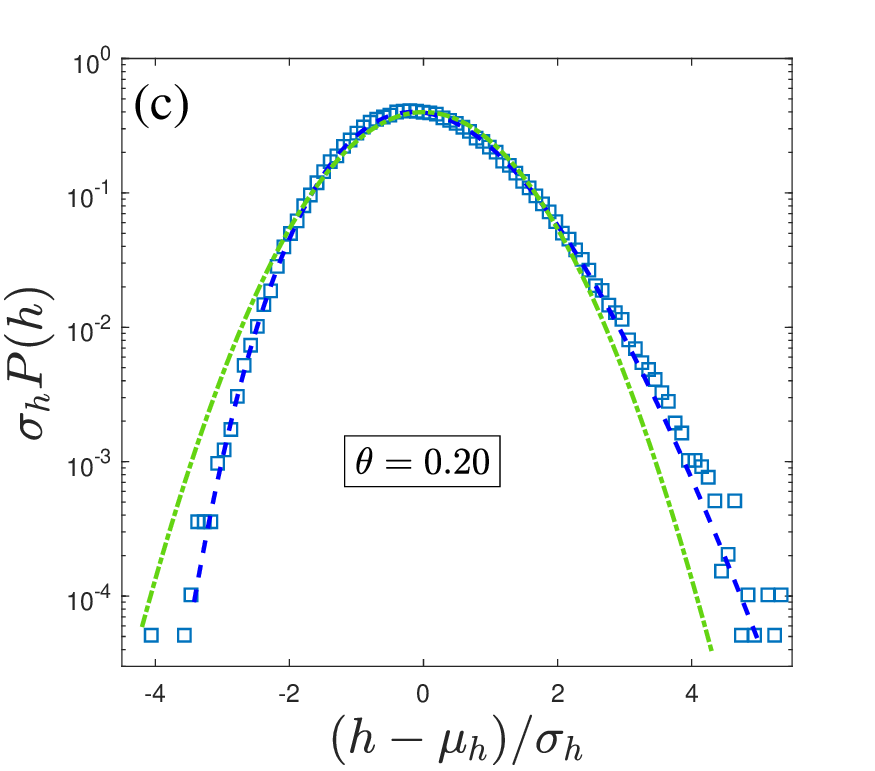}
\end{minipage}
}
\subfigure{
\begin{minipage}[t]{0.40\linewidth}
\centering
\includegraphics[width=2.3in]{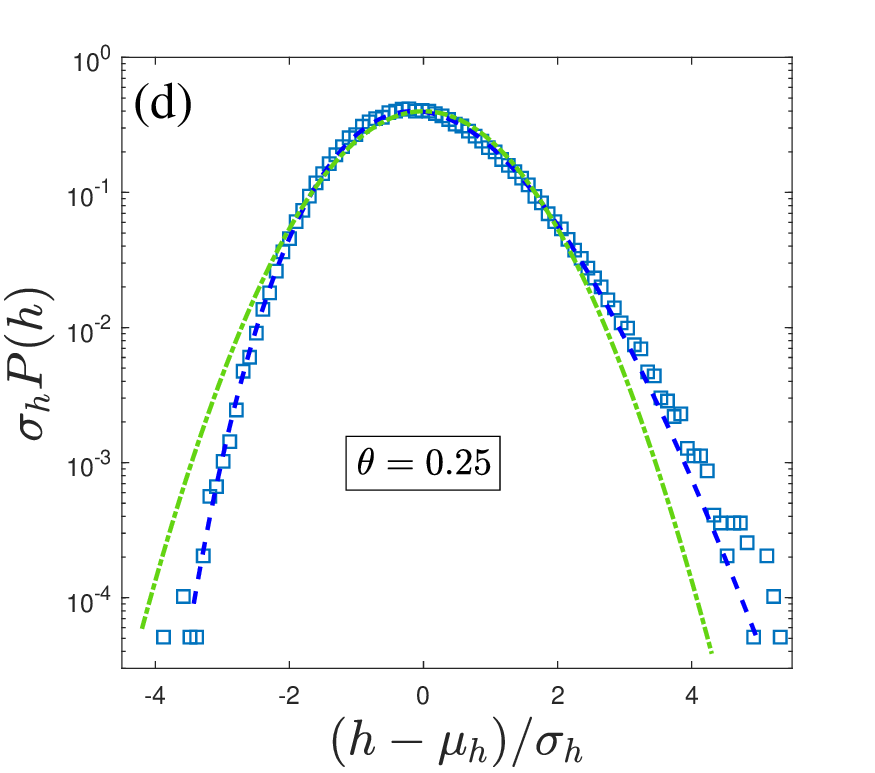}
\end{minipage}
}

\subfigure{
\begin{minipage}[t]{0.40\linewidth}
\centering
\includegraphics[width=2.3in]{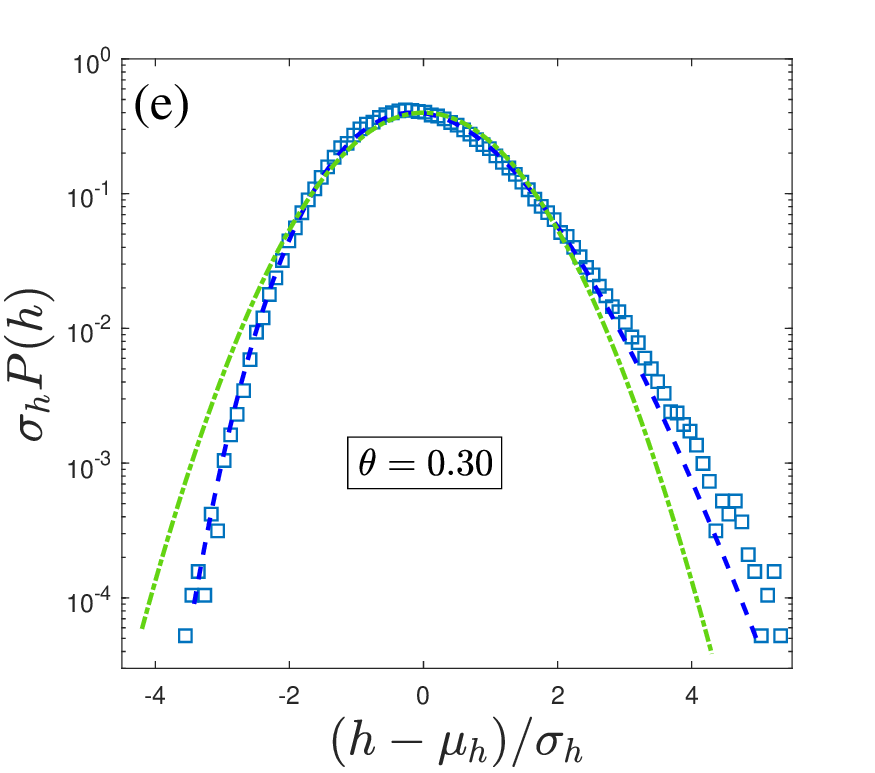}
\end{minipage}
}
\subfigure{
\begin{minipage}[t]{0.40\linewidth}
\centering
\includegraphics[width=2.3in]{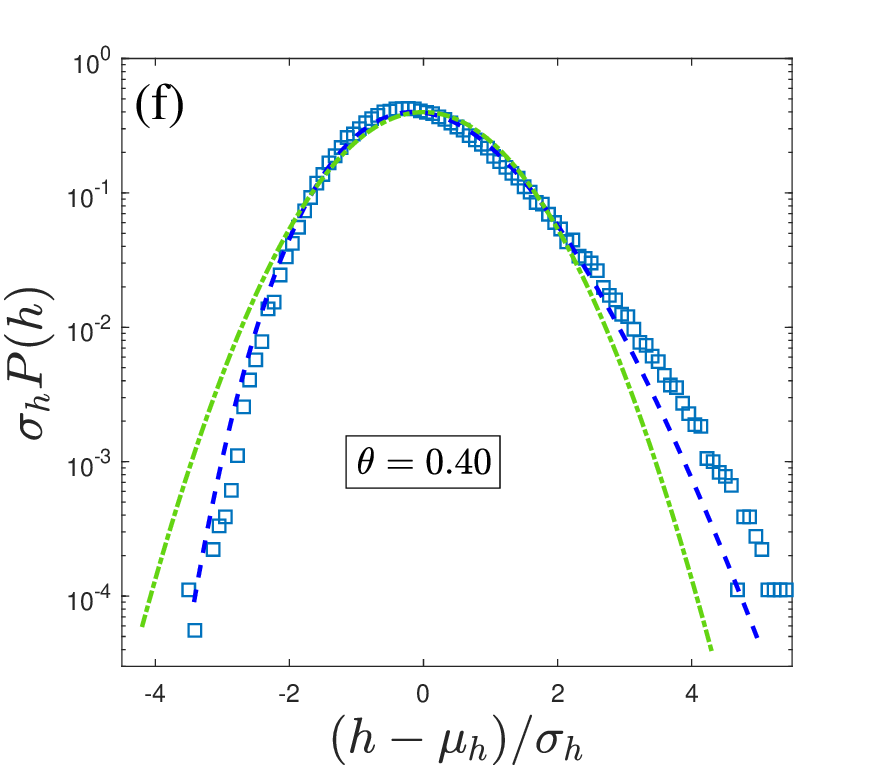}
\end{minipage}
}
\caption{ {The rescaled distributions for the interface height in the temporally correlated KPZ equation with different $\theta$: (a) $\theta = 0.00$, (b) $\theta = 0.10$, (c) $\theta = 0.20$, (d) $\theta = 0.25$, (e) $\theta = 0.30$, (f) $\theta = 0.40$.  Here, $\mu_{h}$ and $\sigma_{h}$ are the mean value and standard deviation of the interface height, respectively. Comparisons with  TW-GOE (blue dash line) and  Gaussian distribution (green dash-dot line) are provided correspondingly.}}
\label{Fig2}
\end{figure}

\begin{figure}[htbp]
\centering
\subfigure{
\begin{minipage}[t]{1\linewidth}
\centering
\includegraphics[width=3.9in]{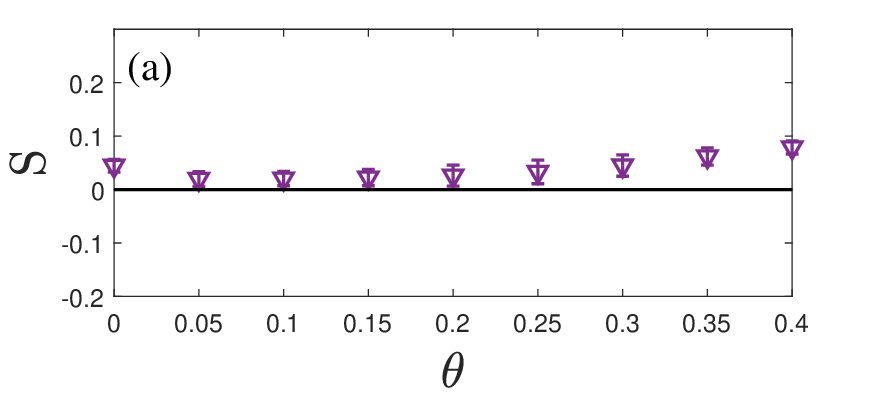}
\end{minipage}
}
\quad
\subfigure{
\begin{minipage}[t]{1\linewidth}
\centering
\includegraphics[width=3.9in]{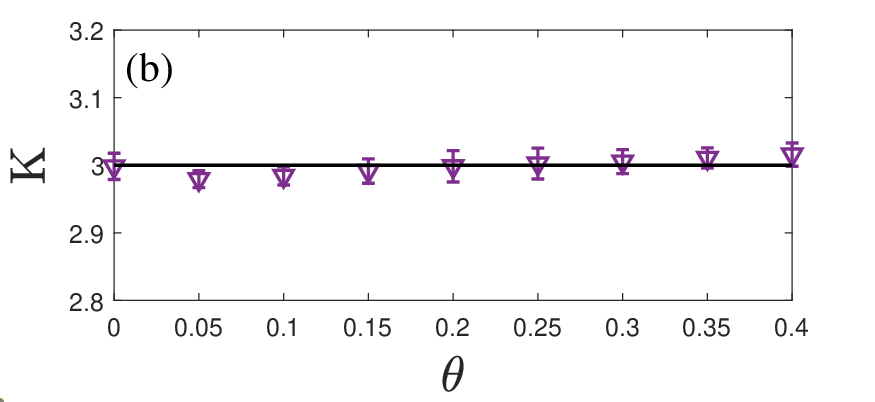}
\end{minipage}
}
\centering
\caption{The estimated values of (a) skewness $S$ and (b) kurtosis $K$ with error bars for probability distributions of the logarithm of the squared interface width for KPZ model with long-range temporal correlation. The Gaussian distribution values (solid lines) are provided for quantitative comparison. }
\label{Fig3}
\end{figure}

\begin{figure}[htbp]
\centering
\subfigure{
\begin{minipage}[t]{1\linewidth}
\centering
\includegraphics[width=3.9in]{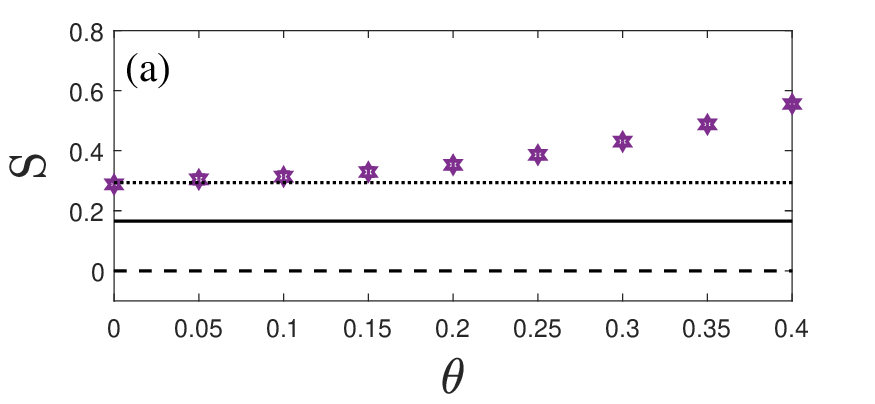}
\end{minipage}
}
\quad
\subfigure{
\begin{minipage}[t]{1\linewidth}
\centering
\includegraphics[width=3.9in]{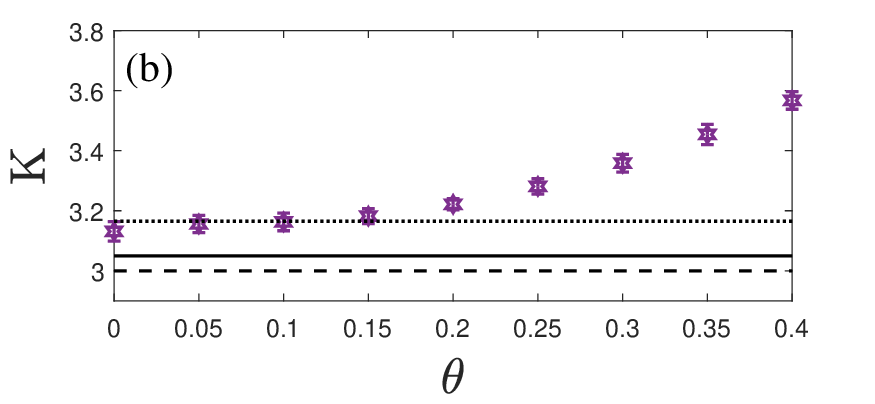}
\end{minipage}
}
\centering
\caption{ {The estimated values of (a) skewness $S$ and (b) kurtosis $K$ with error bars for height distributions for the KPZ model with long-range temporal correlation. The TW-GOE values (dotted lines), TW-GSE values (solid lines) and Gaussian distribution values (dash lines) are provided for quantitative comparison.}}
\label{Fig4}
\end{figure}

\begin{figure}[htbp]
\centering
\subfigure{
\begin{minipage}[t]{0.4\linewidth}
\centering
\includegraphics[width=2.3in]{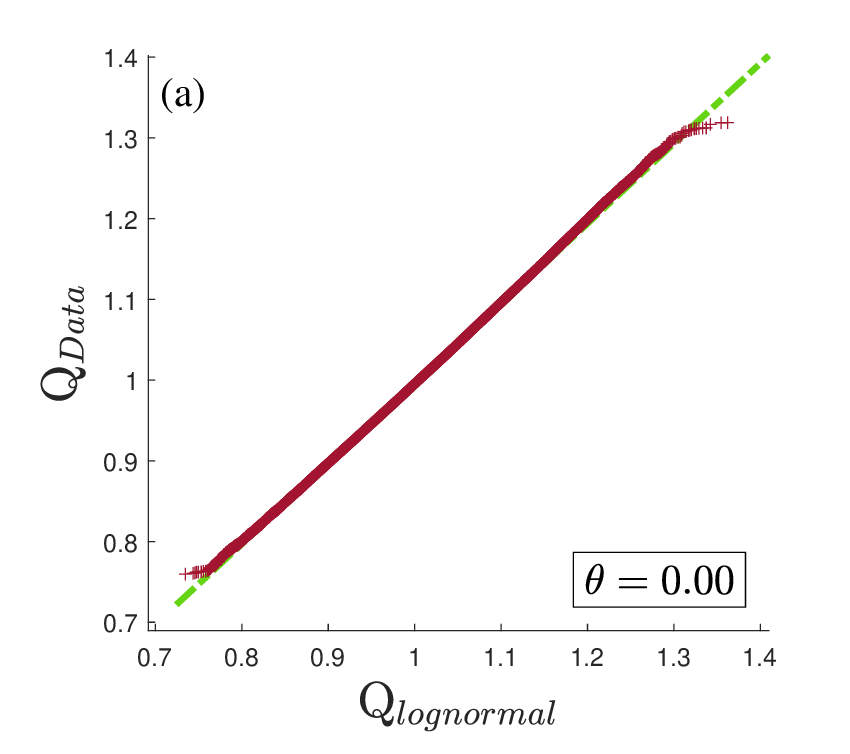}
\end{minipage}
}
\subfigure{
\begin{minipage}[t]{0.4\linewidth}
\centering
\includegraphics[width=2.3in]{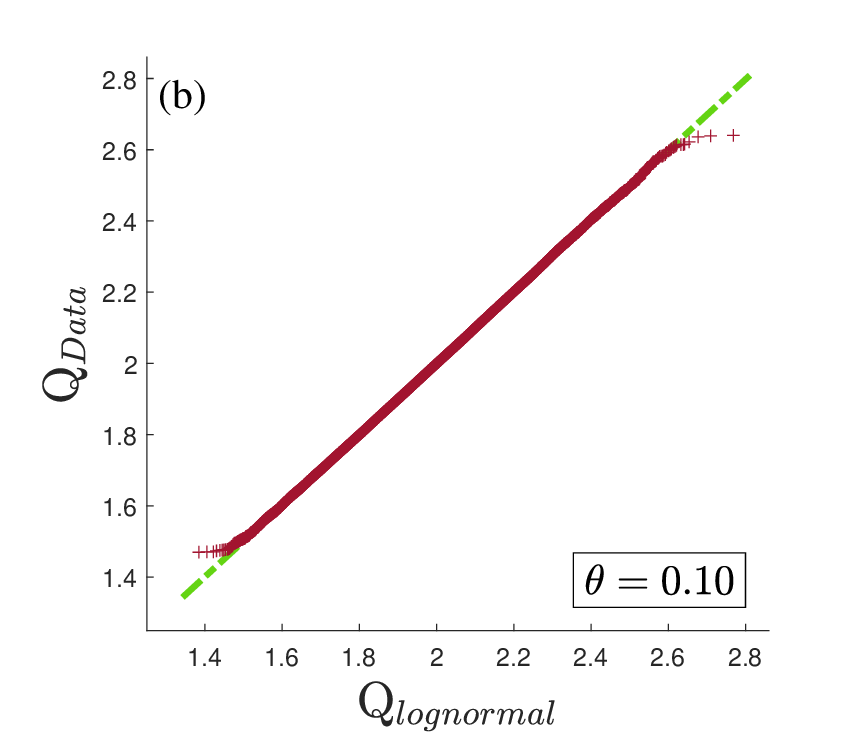}
\end{minipage}
}

\subfigure{
\begin{minipage}[t]{0.4\linewidth}
\centering
\includegraphics[width=2.3in]{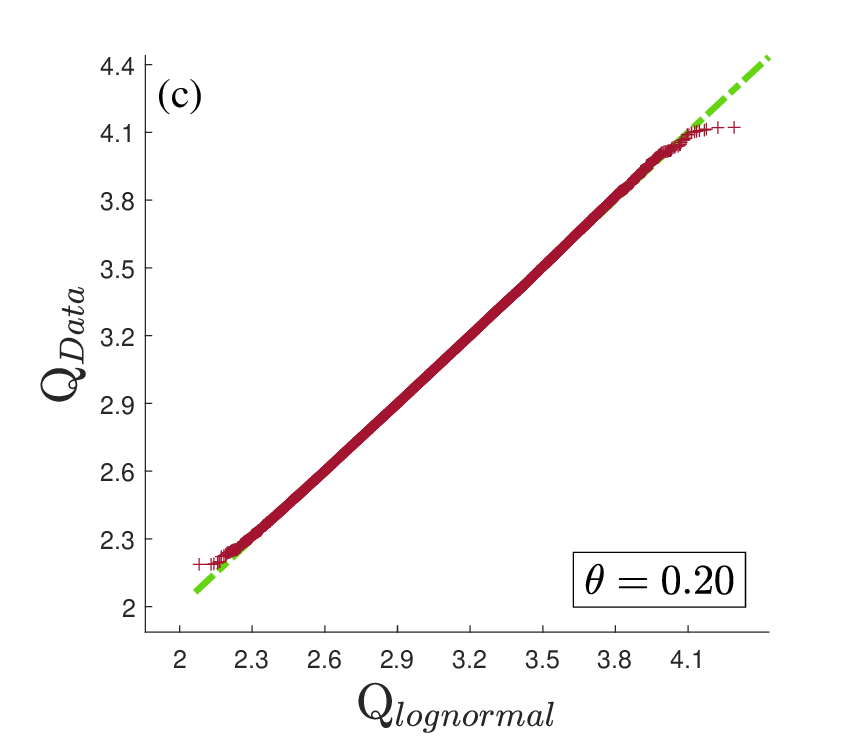}
\end{minipage}
}
\subfigure{
\begin{minipage}[t]{0.4\linewidth}
\centering
\includegraphics[width=2.3in]{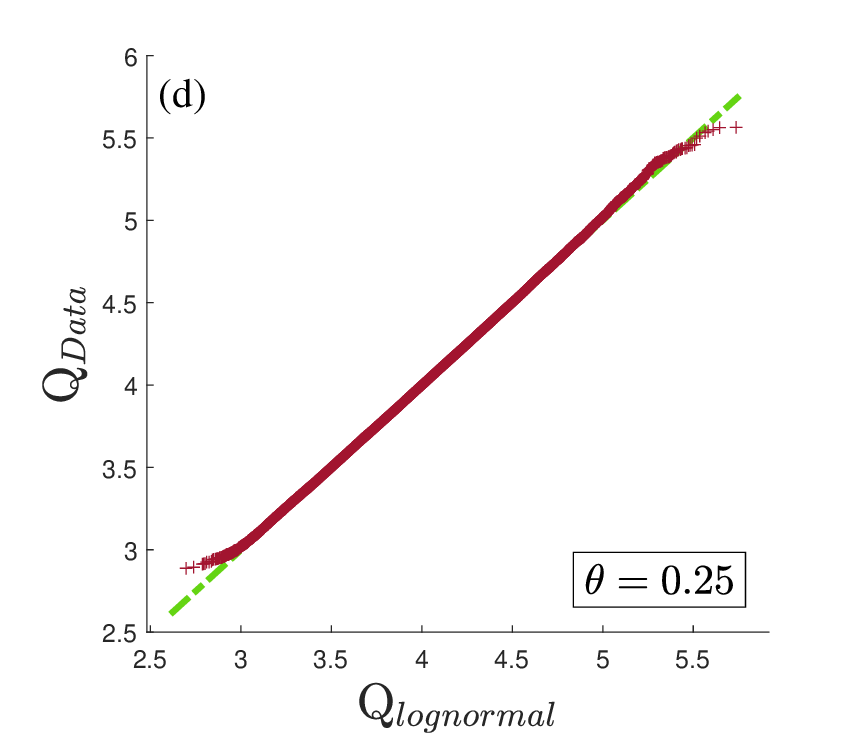}
\end{minipage}
}

\subfigure{
\begin{minipage}[t]{0.4\linewidth}
\centering
\includegraphics[width=2.3in]{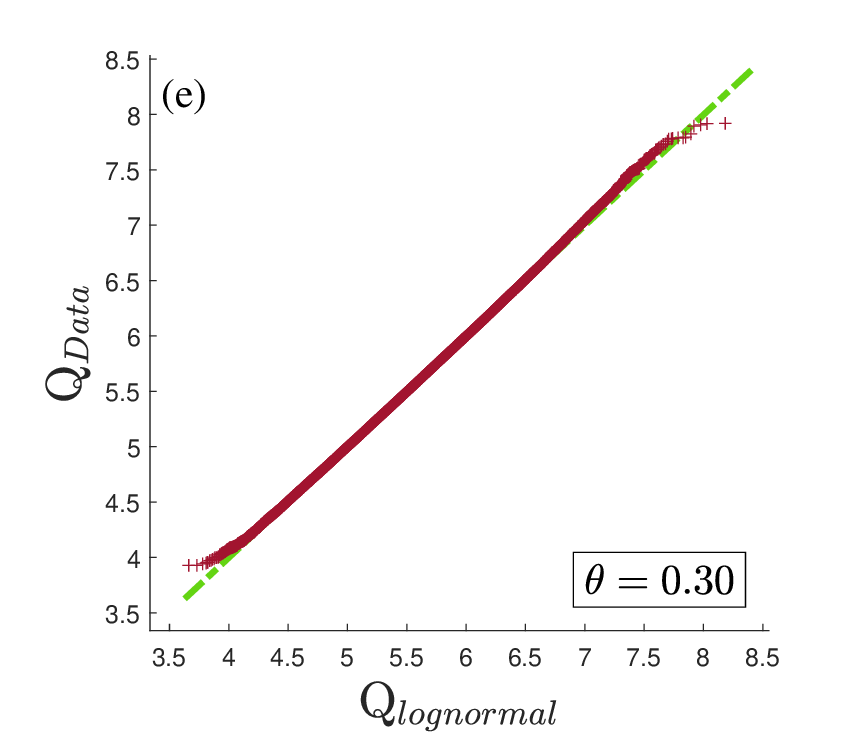}
\end{minipage}
}
\subfigure{
\begin{minipage}[t]{0.4\linewidth}
\centering
\includegraphics[width=2.3in]{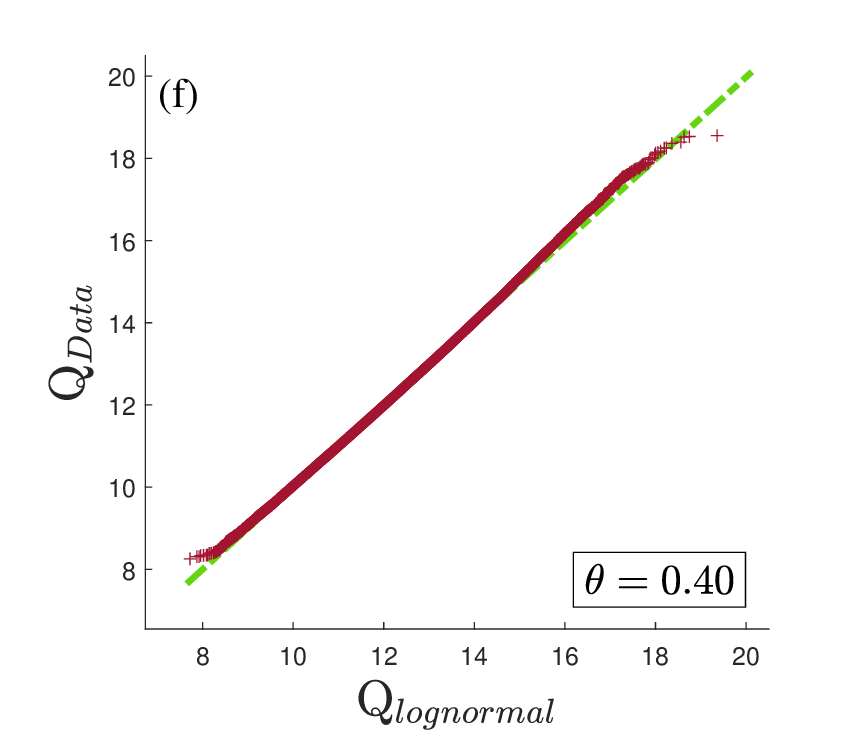}
\end{minipage}
}
\caption{{The Q-Q plots of simulation data of the squared interface width for the temporally correlated KPZ equation versus lognormal distribution with different $\theta$.} }
\label{Fig5}
\end{figure}

In this subsection, we investigate numerically the temporally correlated KPZ system in the (1+1)-dimensions based on the LS numerical scheme. 
{Considering that it is difficult to obtain the probability distributions of the interface width and height in the steady state due to the limitation of our maximum computing power, investigating the universal characteristics of {both $P(W^2)$ and $P(h)$} within the early growth regime becomes the major tasks in our current work.
To study the effects of long-range temporal correlation on probability distribution for kinetic roughening within the early growth regimes, we calculate the time evolution of the growth height and the squared interface width with periodic boundary conditions. Here, system size $L = 4096$ and growth time $t = 300$ are used.  Figures \ref{Fig1} and \ref{Fig2} show the obtained results of $P(W^2)$ and $P(h)$ with different $\theta$,  respectively.} 

{Firstly, our results about $P(W^2)$ indicate that, within the small $\theta$ regimes, the rescaled distribution of the squared interface width fits highly the lognormal distribution. With increasing $\theta$, the changing trench of the squared width distributions is not significant apart from minor deviations in the right tail, as shown in  Fig.\ref{Fig1}. Thus, the temporal correlation has a little effect on the rescaled distributions of the squared interface width. }

{Then, in order to investigate further temporal correlation effects, and give a detailed comparison between $P(W^2)$ and $P(h)$, we also provide $P(h)$ with different $\theta$, as shown in  Fig.\ref{Fig2}.  Similarly, within the small correlated regimes, the height distributions exhibit positive skewness, meanwhile, they have long tails in the right part, and fit generally the Tracy-Widom Gaussian orthogonal ensemble (TW-GOE) distribution. However, with increasing  $\theta$, the changing trend with a longer tail in the right part is more obvious, implying that the long-range temporal correlation has an evident effect on  $P(h)$ within the large correlation regimes. In short, different from the weak dependence of $P(W^2)$ on the temporal correlation exponent, we find that the $P(h)$ exhibits a more evident dependence on long-range temporal correlations with increasing $\theta$. And the changing trends of the height distribution are similar to those of the growth exponents depending on the values of $\theta$ for the temporally correlated KPZ growth \cite{Song2021b,Hu2023}. }

{Figure \ref{Fig3} shows the estimated values of $S$ and $K$ with error bars for the probability distributions of the logarithm of the squared interface width for different $\theta$ at the early growth time, respectively. We find that when $\theta \to 0$, the estimated values of $S$ and $K$ for the logarithm of $W^2$ are approximately equal to the corresponding exact values for the Gaussian distribution values, namely, the probability distributions of the squared interface width approximately fit the lognormal distribution.
With increasing $\theta$ from zero, for the logarithm of $W^2$, $S$ and $K$ also suddenly decline, and then $S$ gradually increases with an obvious trend while $K$  has little change, which also implies that the effect of long-range temporal correlation on the distributions of $W^2$ really exists. Furthermore, one could find that, when $\theta < 0.20$, the estimated values of both $S$ and $K$ approach to the Gaussian values, which means $P(W^2)$ have a similar type and all agree with the lognormal distribution. With increasing $\theta$, the values of $S$ exhibit a little increase in dependence on $\theta$. 
Notably, we also find that there exist weak finite-time effects, which cause the estimated values of $S$ and $K$ to have a little time dependence within the early growth regimes.}

{Similarly, we have also calculated variations of $S$ and $K$  of the interface height with different $\theta$, the estimated results of  $S$ and $K$ are shown in Figs. \ref{Fig4}(a) and \ref{Fig4}(b), respectively. By full comparison, we find that the changing trends of $S$ and $K$ of the height distributions are different from those of the squared width distributions in the presence of temporal correlation. More precisely, when $\theta \to 0$, the height distribution satisfies the TW-GOE distribution, and with increasing  $\theta$, the estimated values of both $S$ and $K$  show monotonically increasing trends to an unknown distribution.
To compare our numerical data with the lognormal distribution from another point, we adopt a quantile-quantile (Q-Q) plot, a method to determine intuitively whether two series of numbers obey the same distribution, to analyze the simulation data mentioned above. The corresponding results of the Q-Q plot are shown in Fig.\ref{Fig5}. Through quantitative comparison of the lognormal distribution (${{\rm{Q}}_{lognormal}}$) with simulation data (${{\rm{Q}}_{Data}}$), we find that $P(W^2)$ generally obey lognormal distribution within the whole temporally correlated regimes,  which exhibit similar changing trends with the estimated results of $S$ and $K$ values in Fig.\ref{Fig3}. However, we should point out that the deviation of the lognormal distribution is very little in the chosen $\theta$ regimes even though we carefully distinguish them. Thus, these results show that long-range temporal correlation does not obviously change the distribution form of $W^2(L,t)$ for this kind of correlated KPZ growth. Meanwhile, based on Q-Q plot, we also observe that $P(h)$ generally obeys TW-GOE distribution within the small temporally correlated regimes, and with increasing  $\theta$, the deviation between the two quantile plots is gradually increasing (not shown). The changing trend of Q-Q plot is agreement with the $\theta$-dependency on $S$ and $K$ values, as shown in Fig.\ref{Fig4}.}

\subsection{The KPZ equation with long-range spatial correlation}

\begin{figure}[htbp]
\centering
\subfigure{
\begin{minipage}[t]{0.4\linewidth}
\centering
\includegraphics[width=2.3in]{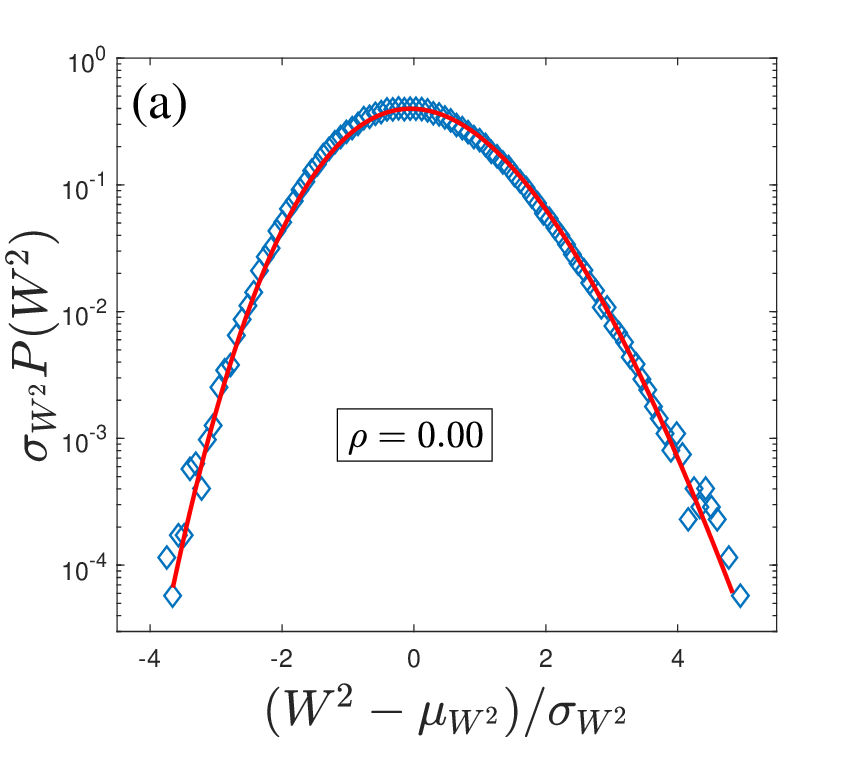}
\end{minipage}
}
\subfigure{
\begin{minipage}[t]{0.4\linewidth}
\centering
\includegraphics[width=2.3in]{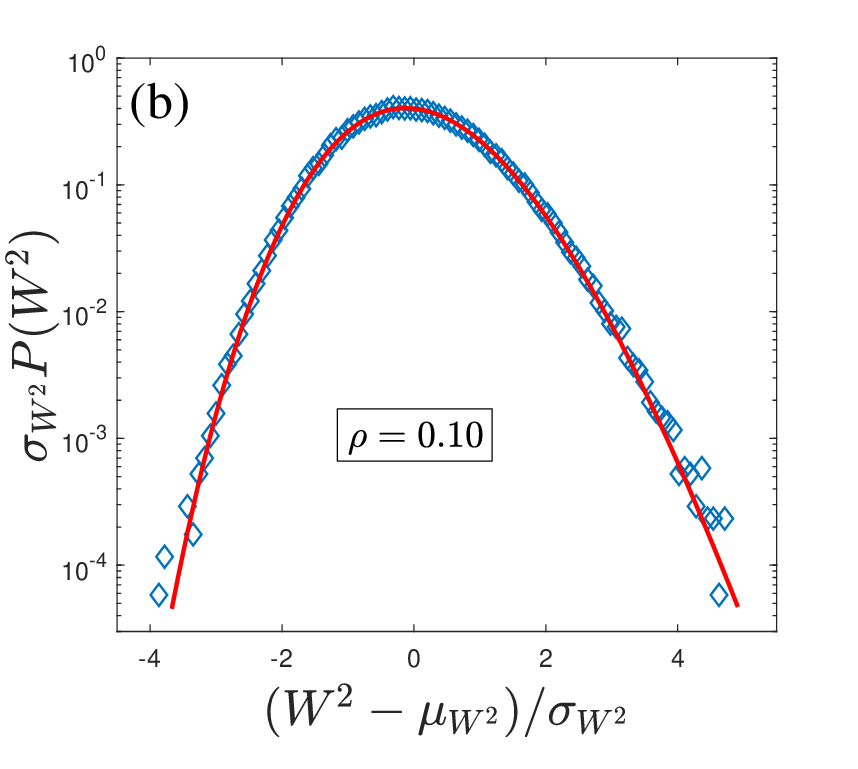}
\end{minipage}
}

\subfigure{
\begin{minipage}[t]{0.4\linewidth}
\centering
\includegraphics[width=2.3in]{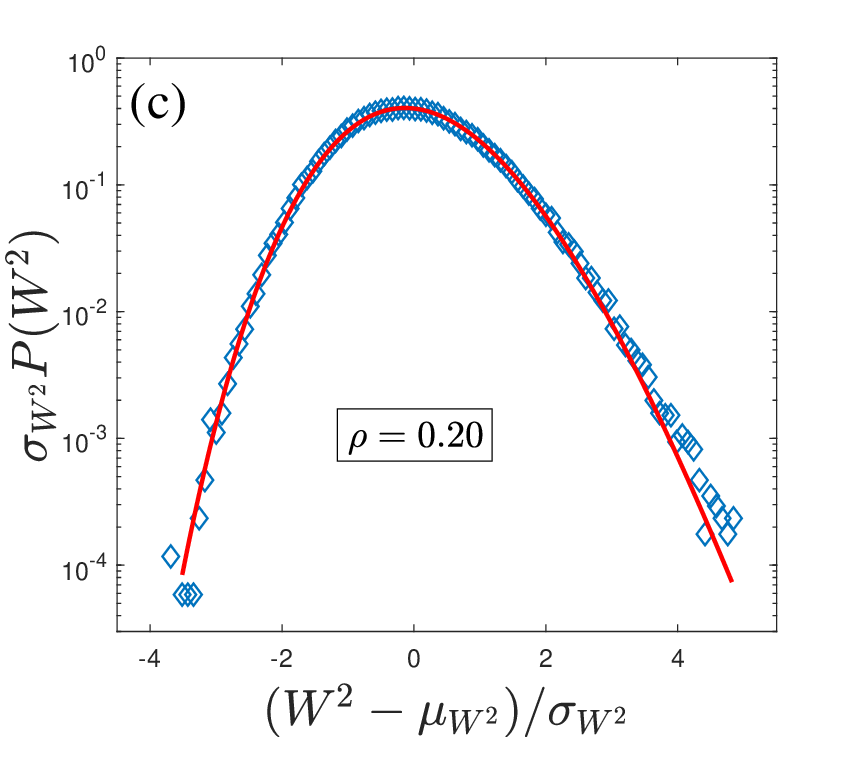}
\end{minipage}
}
\subfigure{
\begin{minipage}[t]{0.4\linewidth}
\centering
\includegraphics[width=2.3in]{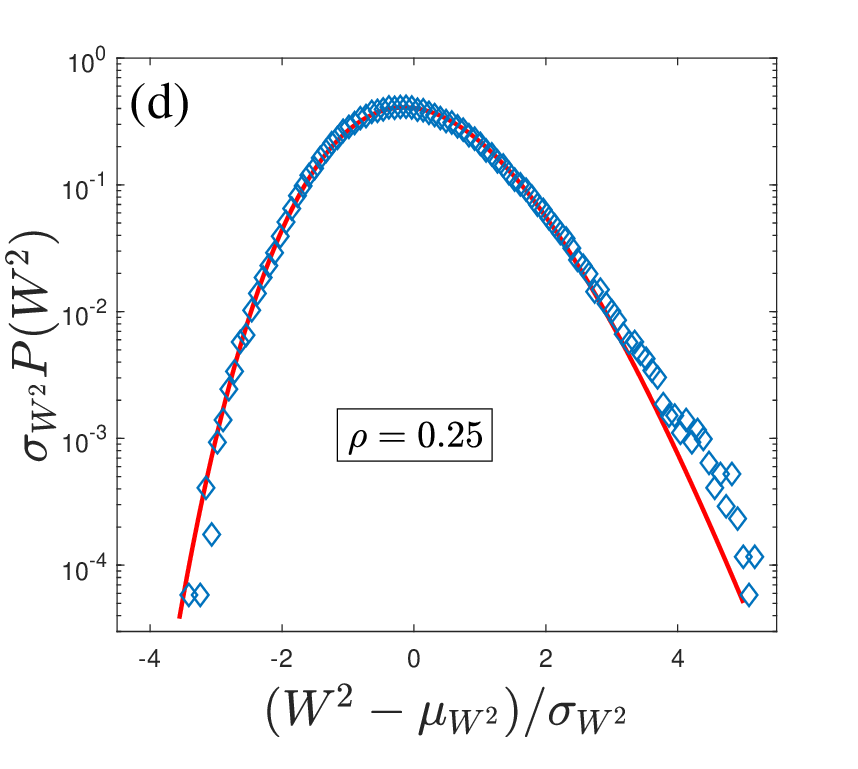}
\end{minipage}
}

\subfigure{
\begin{minipage}[t]{0.4\linewidth}
\centering
\includegraphics[width=2.3in]{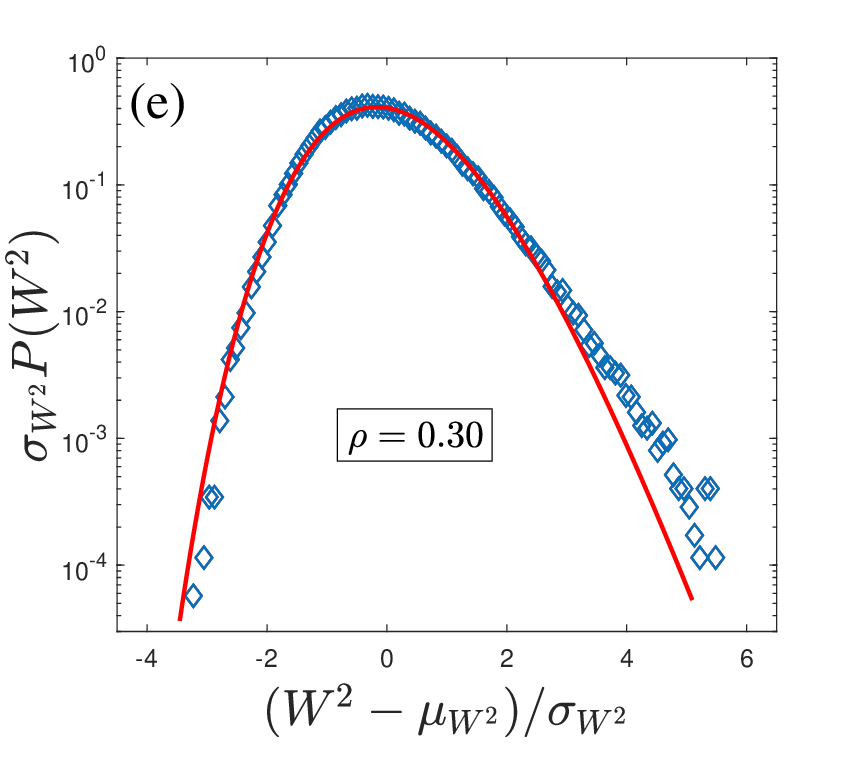}
\end{minipage}
}
\subfigure{
\begin{minipage}[t]{0.4\linewidth}
\centering
\includegraphics[width=2.3in]{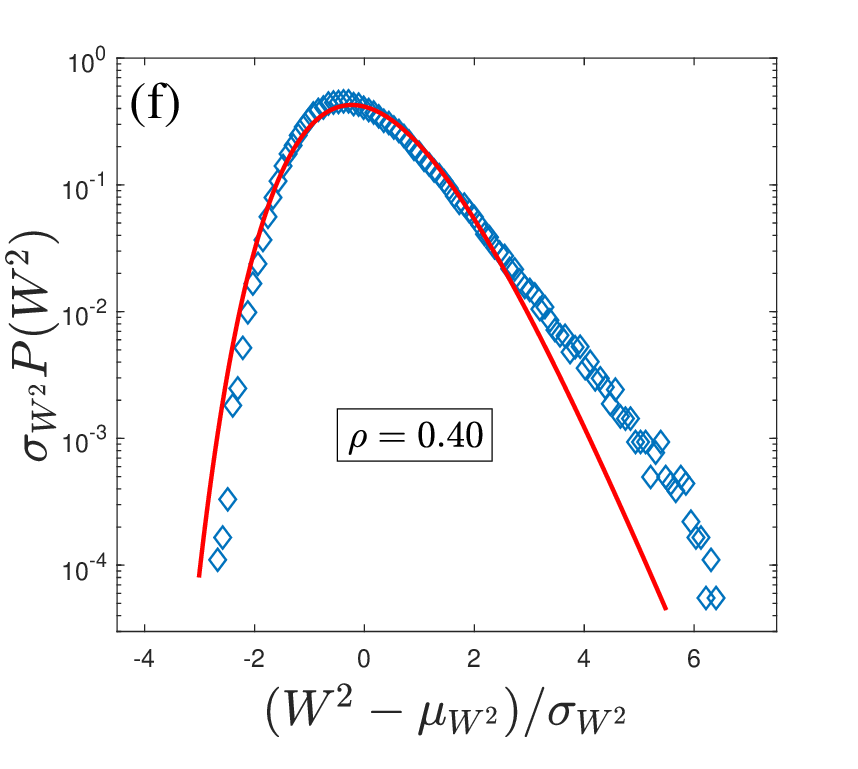}
\end{minipage}
}
\caption{{The rescaled distributions for $W^2(L,t)$ in the spatially correlated KPZ system with different $\rho$: (a) $\rho = 0.00$, (b) $\rho = 0.10$, (c) $\rho = 0.20$, (d) $\rho = 0.25$, (e) $\rho = 0.30$, (f) $\rho = 0.40$. Here, $\mu_{W^2}$ and $\sigma_{W^2}$ represent the mean value and standard deviation of $W^2$, respectively. All data are averaged over $10^5$ independent realizations.  Comparison with lognormal distribution (red solid line) is provided correspondingly.}}
\label{Fig6}
\end{figure}

\begin{figure}[htbp]
\centering
\subfigure{
\begin{minipage}[t]{0.4\linewidth}
\centering
\includegraphics[width=2.3in]{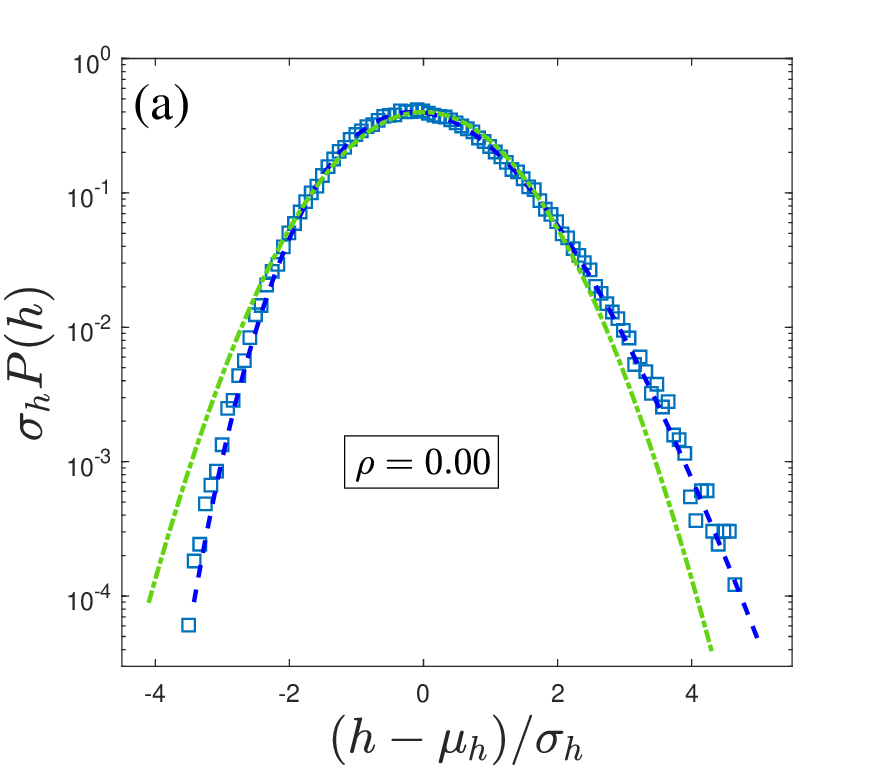}
\end{minipage}
}
\subfigure{
\begin{minipage}[t]{0.4\linewidth}
\centering
\includegraphics[width=2.3in]{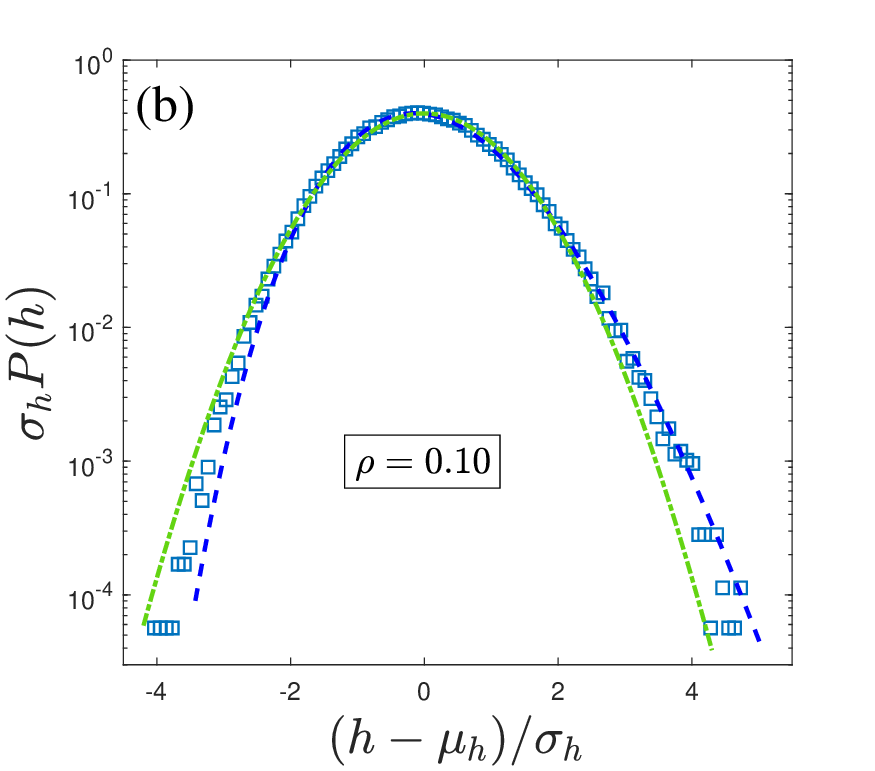}
\end{minipage}
}

\subfigure{
\begin{minipage}[t]{0.4\linewidth}
\centering
\includegraphics[width=2.3in]{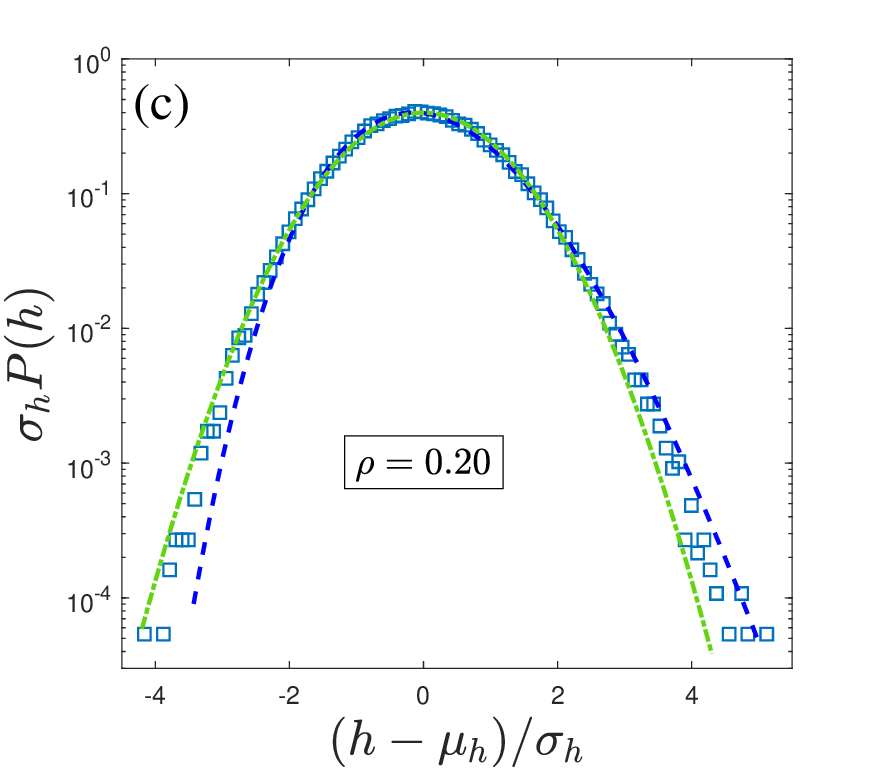}
\end{minipage}
}
\subfigure{
\begin{minipage}[t]{0.4\linewidth}
\centering
\includegraphics[width=2.3in]{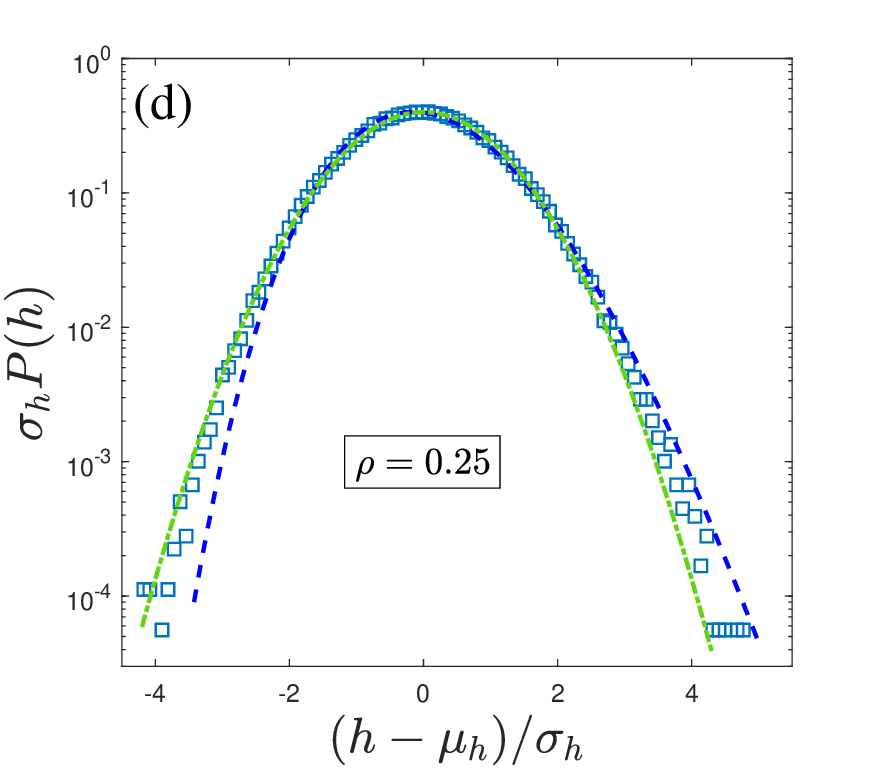}
\end{minipage}
}

\subfigure{
\begin{minipage}[t]{0.4\linewidth}
\centering
\includegraphics[width=2.3in]{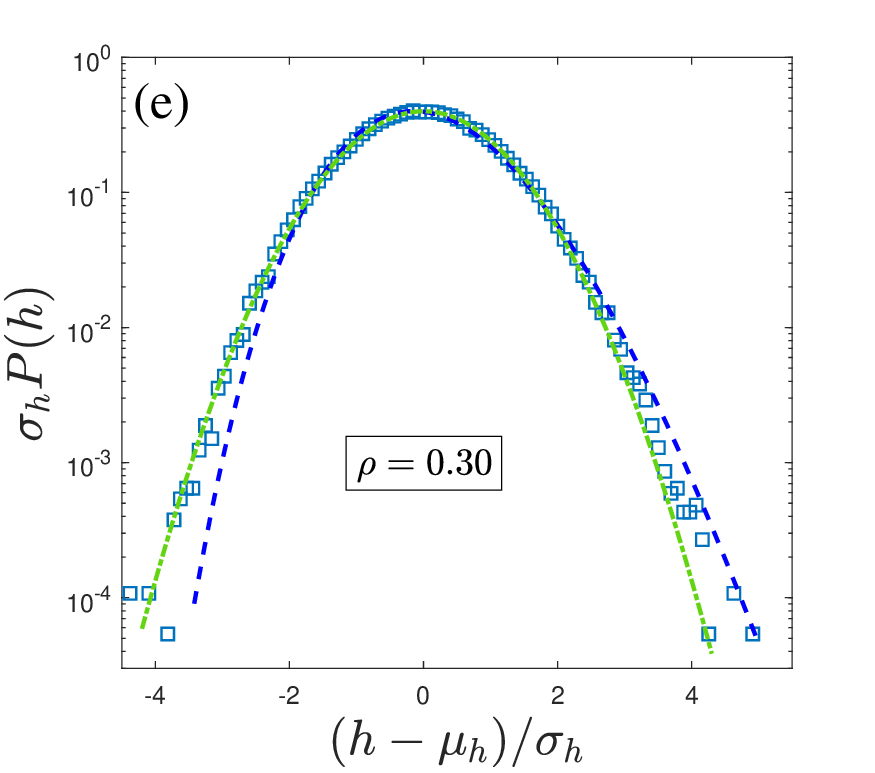}
\end{minipage}
}
\subfigure{
\begin{minipage}[t]{0.4\linewidth}
\centering
\includegraphics[width=2.3in]{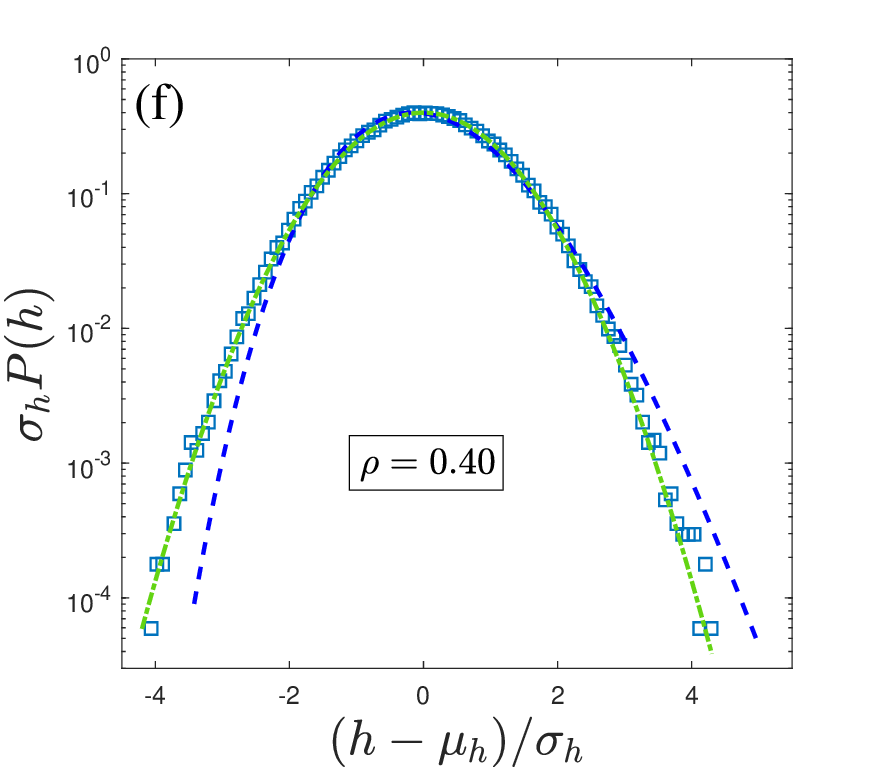}
\end{minipage}
}
\caption{ {The rescaled distributions for the interface height in the spatially correlated KPZ system with different $\rho$. Here $\mu_{h}$ and $\sigma_{h}$ are the mean value and standard deviation of the interface height, respectively. All parameters chosen here are the same as those in Fig.\ref {Fig6}. The suitable distributions are provided for comparison: TW-GOE (blue dash line) and Gaussian distribution (green dash-dot line).}}
\label{Fig7}
\end{figure}

\begin{figure}[htbp]
\centering
\subfigure{
\begin{minipage}[t]{1\linewidth}
\centering
\includegraphics[width=3.9in]{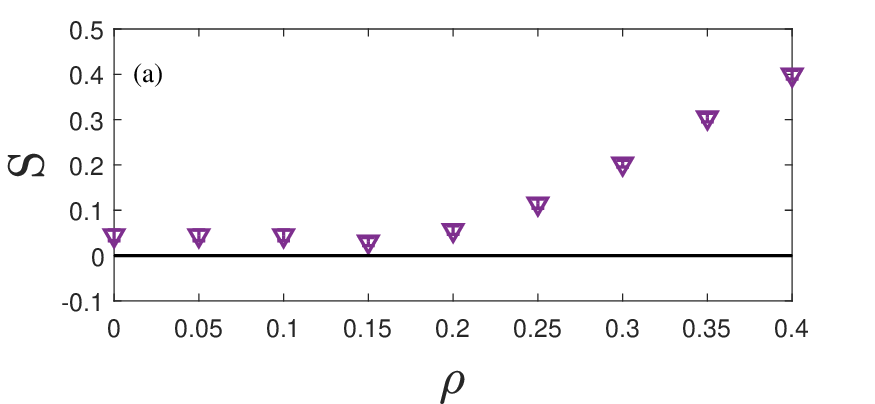}
\end{minipage}
}
\quad
\subfigure{
\begin{minipage}[t]{1\linewidth}
\centering
\includegraphics[width=3.9in]{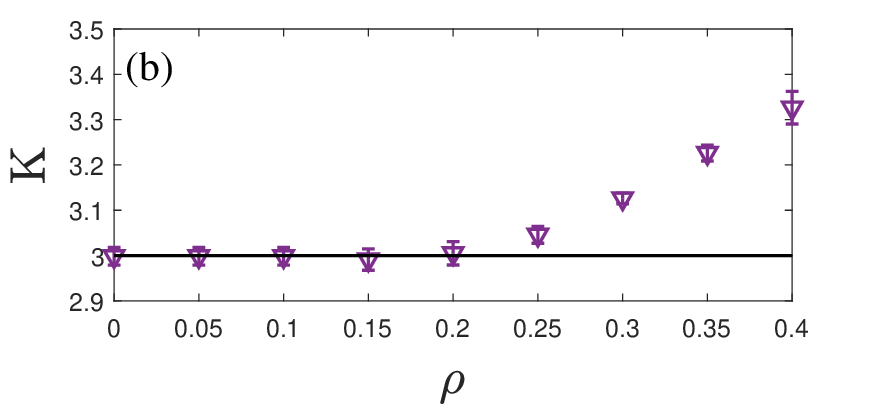}
\end{minipage}
}
\centering
\caption{The estimated values of (a) skewness $S$ and (b) kurtosis $K$ with error bars for {probability distributions of the logarithm of the squared interface width $W^2(L,t)$ (purple downward-pointing triangle) in the spatially correlated KPZ system with different $\rho$,  which compare quantitatively with the Gaussian distribution values (solid lines). } }
\label{Fig8}
\end{figure}

\begin{figure}[htbp]
\centering
\subfigure{
\begin{minipage}[t]{1\linewidth}
\centering
\includegraphics[width=3.9in]{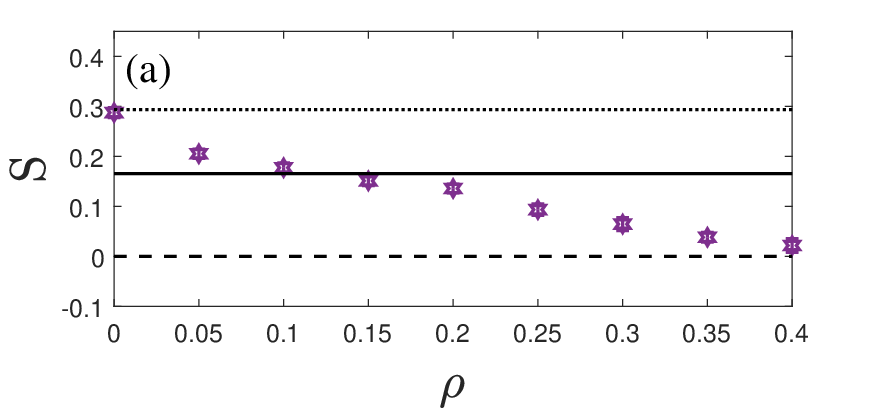}
\end{minipage}
}
\quad
\subfigure{
\begin{minipage}[t]{1\linewidth}
\centering
\includegraphics[width=3.9in]{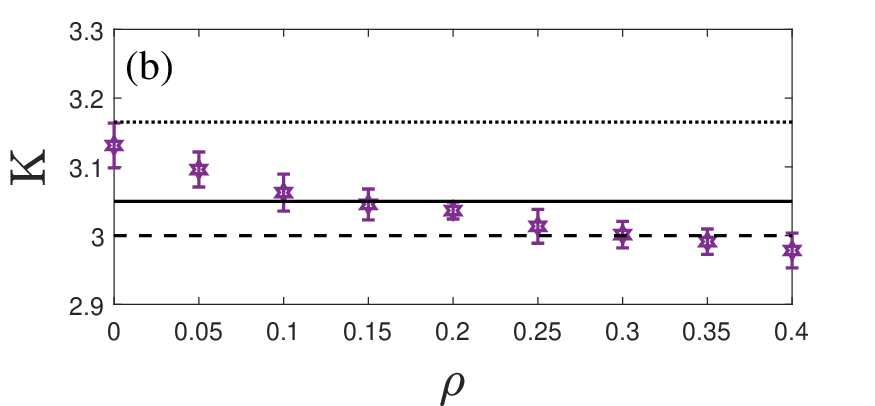}
\end{minipage}
}
\centering
\caption{{The estimated values of (a) skewness $S$ and (b) kurtosis $K$ with error bars for height distributions for the KPZ model with long-range spatial correlation. The TW-GOE values (dotted lines), the  TW-GSE values (solid lines) and the Gaussian distribution values (dash lines) are provided for quantitative comparison.}}
\label{Fig9}
\end{figure}

{To investigate probability distributions for the interface width and height of the spatially correlated KPZ system within the early growth regimes, we perform numerical simulations with different spatial correlation exponent $\rho$. System size $L = 4096$ and growth time $t = 300$ are also used, and numerical results of the interface width distribution are presented in Figs.\ref{Fig6} and \ref{Fig7}. We find that the $P(W^2)$ quantitatively fits lognormal distribution when $\rho \le 0.20$, and with $\rho$ increasing, the transfer tends to other distribution types, which evidently differ from those of the temporal correlatedly KPZ growth.
And when  $\rho \ge 0.25$, the distributions tend to the unknown forms, especially exhibiting relatively large deviation on the right tail with increasing $\rho$.}
In order to provide a detailed comparison between the interface width and height distributions, we also calculate the height distributions with different $\rho$, as shown in  Fig.\ref{Fig8}. Similar to the evident dependence of the interface width distributions within the large spatially correlated regimes, we notice that the {$P(h)$ exhibit a certain dependence on long-range spatial correlations during the chosen $\rho$ regimes.  Notedly, with increasing $\rho$, the left tail gradually deviates from the existing TW-GOE distribution within the small $\theta$ regimes. Interestingly, the changing trends of the rescaled height distributions from TW-GOE to Gaussian forms are also similar to the results for the optimal energy of directed polymers in random media (DPRM) driven by the spatially correlated noise \cite{Chu2016}.}

{To perform more quantitative comparisons, we analyze the variations of $S$ and $K$  for the logarithm of the squared interface width with different $\rho$, the corresponding results are shown in Figs.\ref{Fig8}(a) and \ref{Fig8}(b), respectively. As a special case, when $\rho \to 0$, the obtained values of $S$ and $K$ for the logarithm of $W^2(L,t)$ equal approximately the Gaussian distribution values accordingly, as same as the counterpart of the temporal correlation. We find that, as $\rho$ increases, $S$ and $K$ for $W^2(L,t)$ exhibit similar changing trends.} 
{More specifically, both $S$ and $K$ vary very slowly with $\rho$ increasing within the small $\rho$ regimes, which indicates that the distribution form is not significantly affected by long-range spatially correlated noise. However, when $\rho > 0.20$, the estimated values of $S$ and $K$ appear to increase with $\rho$, the distributions become more right-skewed, and have more fat tails.} 
These results show that long-range spatial correlation could affect gradually the distribution form within the large correlated regimes. Similar to long-range temporal correlation, we also notice that finite-time effects lead to $S$ and $K$ having weak time dependence within the early growth regimes. These characteristics of  $S$ and $K$ are also consistent with our numerical results of distributions in Figs. \ref{Fig6} and \ref{Fig7}.  
Similarly, we have also calculated variations of $S$ and $K$  of interface height with different $\rho$, the estimated results of  $S$ and $K$ are shown in Figs. \ref{Fig9}(a) and \ref{Fig9}(b), respectively. By comparison, one can observe that the changing trends of $S$ and $K$ of interface height distributions are significantly different from those of the squared interface width distributions in the presence of spatial correlation.
{Firstly, with the increase of $\rho$, the estimated values of both $S$ and $K$  for the height distribution exhibit continuous decreasing trends, but the values of the squared width distribution increase gradually. Secondly, the estimated values of both $S$ and $K$ of $P(h)$ approach to the values of the GOE distribution when $\rho \to 0$, and with increasing $\rho$, the estimated values of $S$ for the interface height distribution have a continuous crossover from the GOE distribution to the Gaussian case.  
By comparing these values of $S$ and $K$ mentioned above with those from the DPRM with spatially correlated noise \cite{Chu2016}, we find that there exist similar variation trends with our estimated values, that is, the rescaled distributions for the optimal energy of DPRM  shift smoothly from TW-GOE to Gaussian forms with increasing $\rho$ from $0$ to $1/2$. These also indicate that,  considered that DPRM is believed to belong to KPZ universality class when the uncorrelated noise is introduced, the free energy of DPRM is still equivalent to the interface height of direct simulating KPZ equation in the presence of long-range spatial correlations. }

\begin{figure}[htbp]
\centering
\subfigure{
\begin{minipage}[t]{0.4\linewidth}
\centering
\includegraphics[width=2.3in]{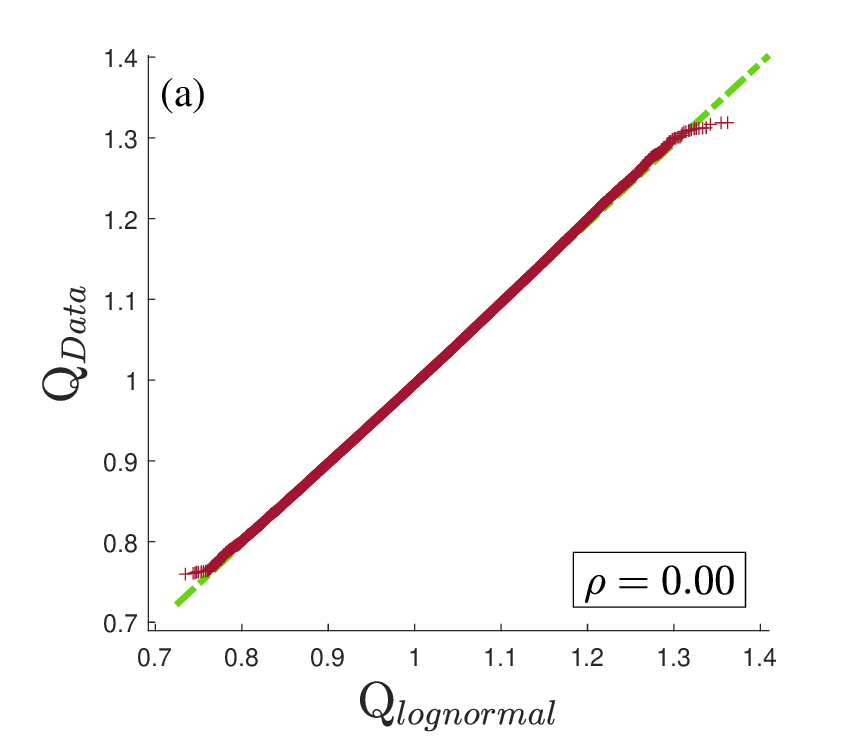}
\end{minipage}
}
\subfigure{
\begin{minipage}[t]{0.4\linewidth}
\centering
\includegraphics[width=2.3in]{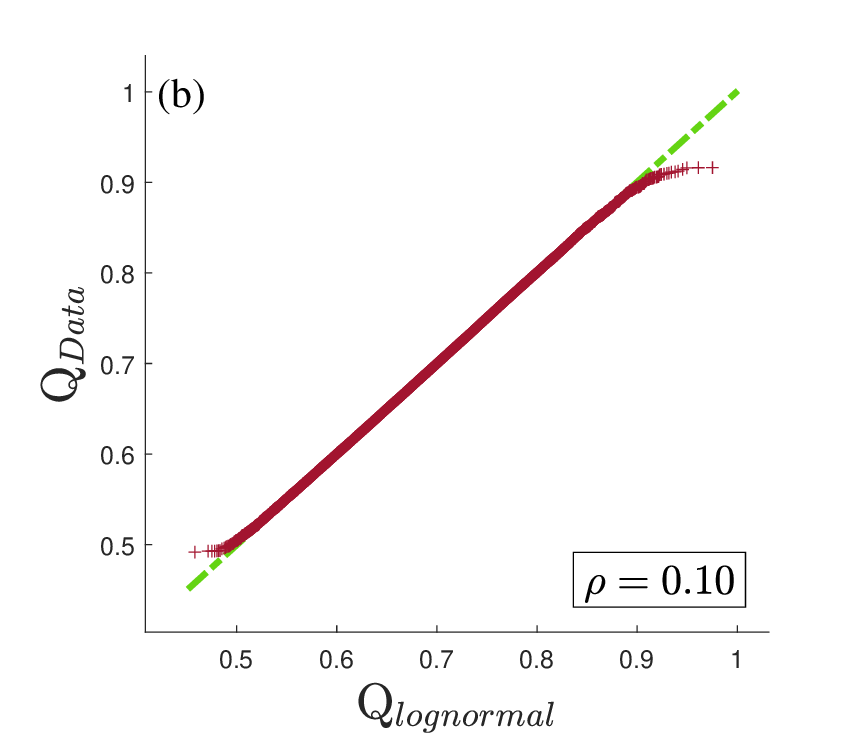}
\end{minipage}
}

\subfigure{
\begin{minipage}[t]{0.4\linewidth}
\centering
\includegraphics[width=2.3in]{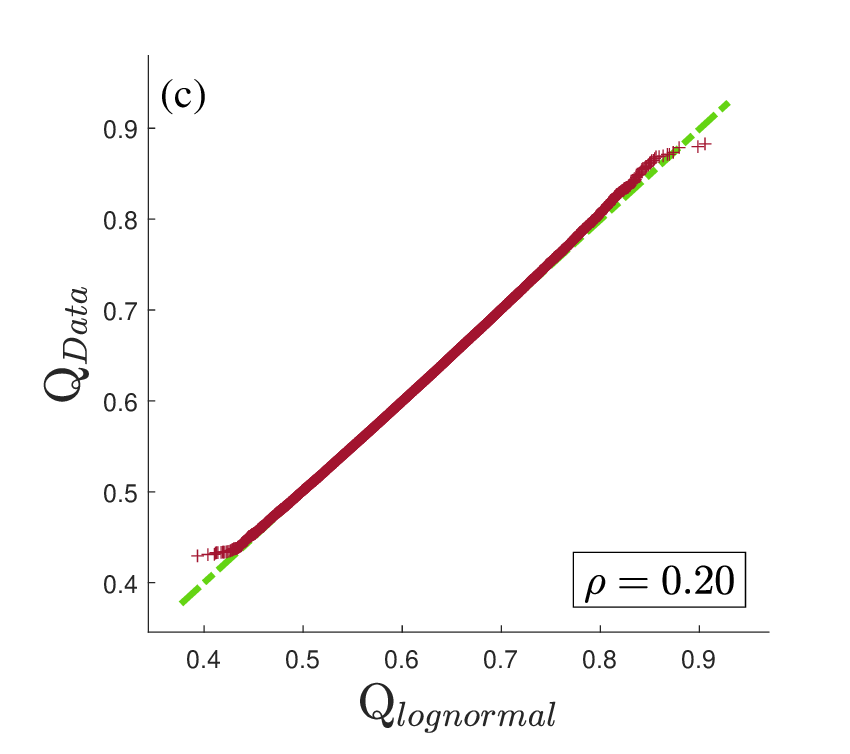}
\end{minipage}
}
\subfigure{
\begin{minipage}[t]{0.4\linewidth}
\centering
\includegraphics[width=2.3in]{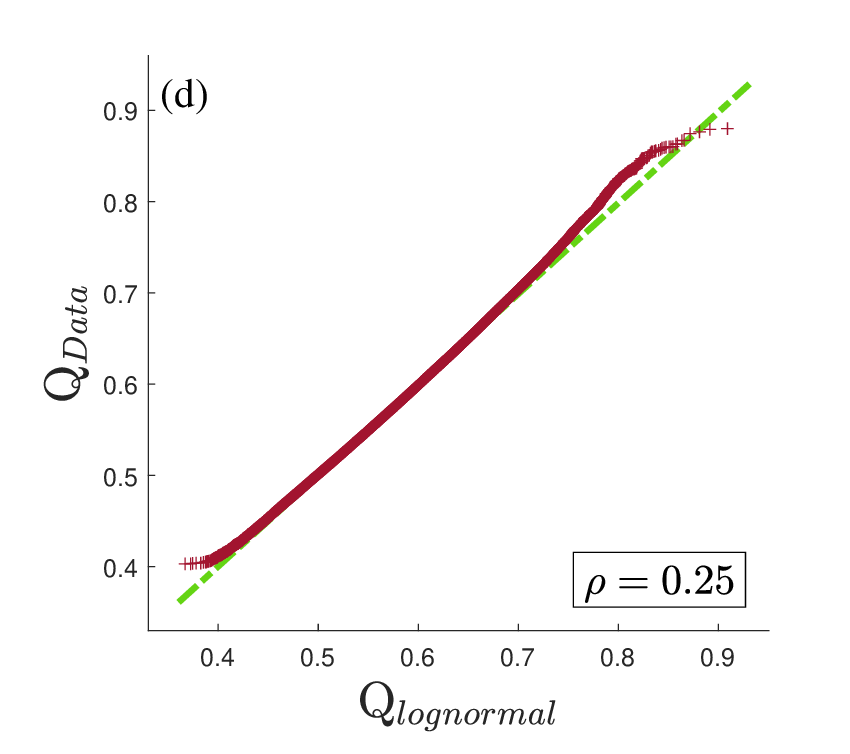}
\end{minipage}
}

\subfigure{
\begin{minipage}[t]{0.4\linewidth}
\centering
\includegraphics[width=2.3in]{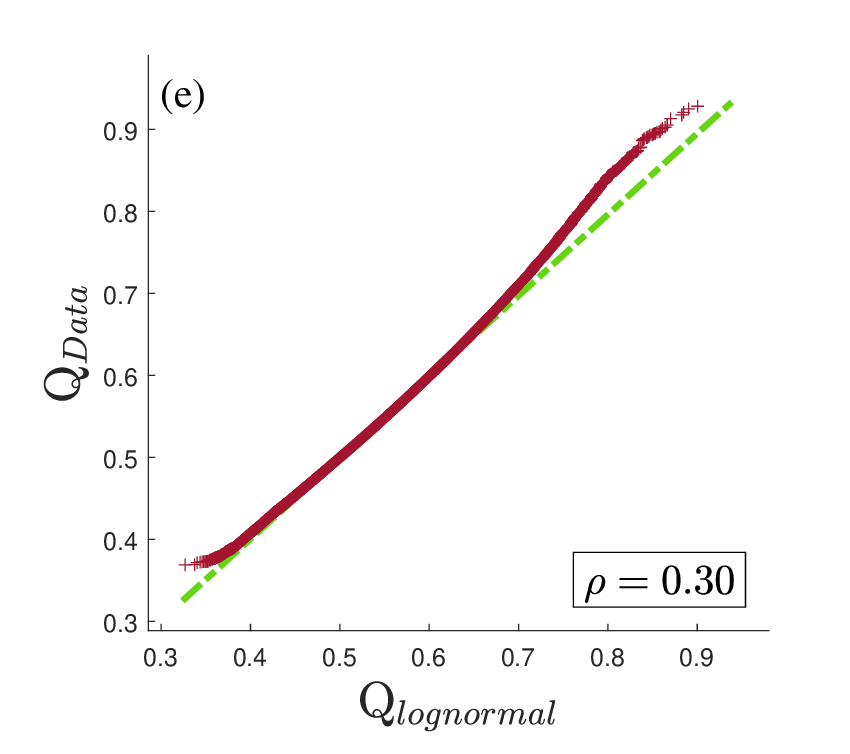}
\end{minipage}
}
\subfigure{
\begin{minipage}[t]{0.4\linewidth}
\centering
\includegraphics[width=2.3in]{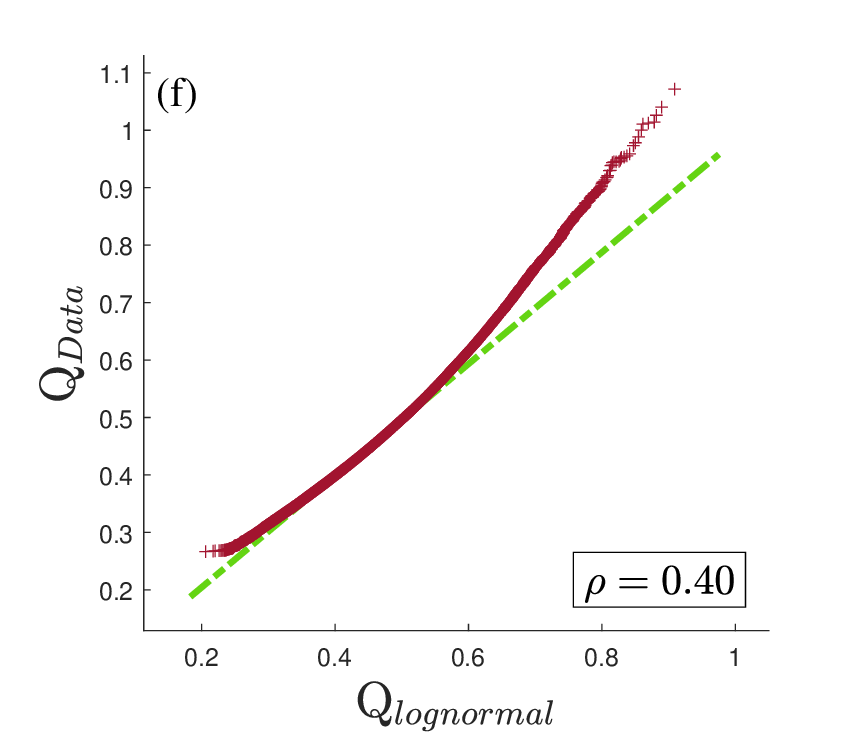}
\end{minipage}
}
\caption{The Q-Q plots of {simulation data of the squared interface width for the spatially correlated KPZ equation versus lognormal distribution with different $\rho$.}}
\label{Fig10}
\end{figure}

{As another independent analytical method, we also compare simulation data of the squared interface width with the lognormal distribution using the Q-Q plot method, as shown in Fig.\ref{Fig10}. These plots for different $\rho$ show similar results with the predictions based on the rescaled width distributions and the corresponding values of $S$ and $K$ mentioned above.  According to a comparison of the quantile of fitting distribution and simulation data, we find that the simulation data fits well with the lognormal distribution within the small spatially correlated regimes. However, there exists an evident discrepancy in the Q-Q plot for the large $\rho$, which implies that simulation data are not completely in agreement with  the lognormal distribution in this correlated regime. Thus, as $\rho$ increases, the squared interface width distributions become more asymmetric, steep, and fat-tailed, it is not very clear which distribution form could obey with the system entering into the strong spatially correlated regime.}

\section{Conclusions}

In summary, we have performed extensive numerical investigations {on probability distributions for kinetic roughening in the (1+1)-dimensional KPZ system with long-range temporally and spatially correlated noises within the early growth regimes. Our results show that long-range temporal and spatial correlations could affect probability distributions of the squared interface width and interface height to a certain extent, and the nontrivial effects of long-range temporal and spatial correlations are obviously different. In the presence of long-range correlations, the correlated KPZ growth systems belong to two different types of continuously changing universality classes, which depend evidently on temporal or spatial correlation exponents. }

For the temporally correlated KPZ equation, probability distributions of {the squared interface width have little effects by temporally correlated noise and are in quantitative agreement with the lognormal distribution within the temporally correlated regimes. And the Q-Q plots further confirm the results, and the variations of $S$ and $K$ also find that the distribution form has a small change with the increase of  $\theta$. However, the height distributions exhibit obvious dependence on long-range temporal correlations, and the changing trends become more evident with increasing $\theta$.
For the spatially correlated KPZ system, our results show that the distribution forms of the squared interface width  are not significantly affected by long-range spatially correlated noise for the small correlated regimes. More precisely, the width distributions of $W^2(L,t)$ could be described by lognormal distribution. With increasing the spatial correlation, the distribution forms of $W^2(L,t)$ evidently depends on $\rho$. And variations of $S$ and $K$ with $\rho$ also show a strong and positive dependence for the large spatially correlated regime. Moreover, the distributions become more skew and thin with increasing  $\rho$. Finally, the distribution of $W^2(L,t)$ tends to an unknown distribution, especially on the right tail, which implies that the spatial correlate KPZ equation smoothly tends to another universality class when $\rho$ increases gradually. However, our results show that there exists the negative dependence on $\rho$ for the height distributions within the whole spatially correlated regimes, that is, the estimated values of both $S$ and $K$ of $P(h)$ decrease gradually with increasing $\rho$. And correspondingly, the height distributions crossover continuously from the TW-GOE distribution to the Gaussian case.}

 In addition, it is important to note that, long-range temporal correlation is less dramatic than long-range spatial correlation in the interface width distributions of the growing surface, especially within the large correlated regions, yet the estimated values of scaling exponents and the changing trends of interface height distributions show stronger dependence on temporal correlation than spatial correlation. Therefore, it is still a lack of a direct link between the estimated values of the scaling exponents and the distribution forms of the interface width in the presence of the long-range temporal and spatial correlations.

\section* {Acknowledgements}
This work is supported by Key Academic Discipline Project of China University of Mining and Technology (CUMT) under Grant No. 2022WLXK04, and Undergraduate Training Program for Innovation and Entrepreneurship of CUMT under Grant No. 202110290059Z.








\biboptions{sort&compress}
\bibliographystyle{elsarticle-num} 
\bibliography{manuscript}
\end{document}